# OCTA-500: A Retinal Dataset for Optical Coherence Tomography Angiography Study

Mingchao Li, Kun Huang, Qiuzhuo Xu, Jiadong Yang, Yuhan Zhang, Zexuan Ji, Keren Xie, Songtao Yuan, Qinghuai Liu, and Qiang Chen

**Abstract**—Optical coherence tomography angiography (OCTA) is a novel imaging modality that has been widely utilized in ophthalmology and neuroscience studies to observe retinal vessels and microvascular systems. However, publicly available OCTA datasets remain scarce. In this paper, we introduce the largest and most comprehensive OCTA dataset dubbed OCTA-500, which contains OCTA imaging under two fields of view (FOVs) from 500 subjects. The dataset provides rich images and annotations including two modalities (OCT/OCTA volumes), six types of projections, four types of text labels (age / gender / eye / disease) and seven types of segmentation labels (large vessel/capillary/artery/vein/2D FAZ/3D FAZ/retinal layers). Then, we propose a multi-object segmentation task called CAVF, which integrates capillary segmentation, artery segmentation, vein segmentation, and FAZ segmentation under a unified framework. In addition, we optimize the 3D-to-2D image projection network (IPN) to IPN-V2 to serve as one of the segmentation baselines. Experimental results demonstrate that IPN-V2 achieves an ~10% mIoU improvement over IPN on CAVF task. Finally, we further study the impact of several dataset characteristics: the training set size, the model input (OCT/OCTA, 3D volume/2D projection), the baseline networks, and the diseases. The dataset and code are publicly available at: https://ieee-dataport.org/open-access/octa-500.

**Keywords**—Medical image dataset, retina, OCTA, segmentation

## 1 Introduction

Optical coherence tomography (OCT) is one of the most significant advances in retinal imaging, as it noninvasively captures 3D structural data of the retina with micron-level resolution (Huang et al., 1991). OCT is widely used to observe the cross-sectional structure of the retina and monitor fluid leakage (Sakata et al., 2009). However, the limitation of OCT is that it cannot directly provide blood flow information. Building on the OCT platform, OCT angiography (OCTA) has been developed as a new useful imaging modality for providing functional information on retinal blood vessels and microvascular systems.

OCTA measures the amplitude and delay of reflected or backscattered light in an interferometrical manner to acquire retinal angiography volume (Kashani et al., 2017). The OCTA volumetric data can be projected from different retinal layers to enable separate visualization of the corresponding retinal plexuses. This unique observation perspective improves our understanding of the pathophysiology of retinal vasculature. Since its first commercial product in 2014, OCTA has quickly demonstrated its excellence in the clinical application of age-related macular degeneration, diabetic retinopathy, choroidal neovascularization, glaucoma, and other eye diseases (Kashani et al., 2017; Spaide et al., 2018; Laíns et al., 2021). More recently, OCTA has also been used to study the functional changes in retinal blood flow in neurological diseases, such as Alzheimer's disease (Jiang et al., 2018; Zabel et al., 2019), Parkinson's disease (Kwapong et al., 2018; Robbin et al., 2022), and Huntington's disease (Maio et al., 2021).

Quantitative OCTA analysis of retinal vasculature is essential to standardize objective interpretations of clinical outcomes (Yao et al., 2020). Quantitative indicators, including vessel density, vessel diameter index, vessel length fraction, vessel fractal dimension, foveal avascular zone (FAZ) area, and foveal avascular zone perimeter have been established for objective OCTA assessment (Wang et al., 2021). To calculate these indicators, it is necessary to segment the vascular structures in OCTA images. OCTA segmentation technology has achieved encouraging advances in recent years (Meiburger et al., 2021), especially in vessel segmentation (Stefan and Lee, 2020; Li et al., 2020a; Giarratano et al., 2020; Mou et al., 2021) and FAZ segmentation (Li et al., 2020a; Díaz et al., 2019; Guo et al., 2019a), also due to technological advances, especially deep learning. Despite these achievements, the quantification method of OCTA is still in development. With the emergence of some new discoveries, the quantification methods are still being improved. For example, differential artery-vein analysis has recently been demonstrated to improve OCTA performance for objective detection and classification of eye disease (Alam et al., 2019; Alam et al., 2018), thus making segmentation of arteries and veins in OCTA images a new research direction (Alam et al., 2020). More recently, the FAZ volume was proposed to depict the 3D structure of the FAZ (Xu et al., 2021), so 3D FAZ segmentation is also a new focus (Xu et al., 2022).

Datasets play a crucial role in computer vision research (Deng et al., 2009). Over the past two decades, we have witnessed tremendous advancements in fundus image research

- *M. Li, K. Huang, Q. Xu, J. Yang, Y. Zhang and Z. Ji is with the School of Computer Science and Engineering, Nanjing University of Science and Technology, Nanjing 210094, China.*
  *E-mail: [chaosli, huangkun, xuqiuzhuo, yangjiadong, zhangyuhan, jizexuan] @njust.edu.cn.*
- *K. Xie, S. Yuan and Q. Liu is with Department of Ophthalmology, The First Affiliated Hospital with Nanjing Medical University, Nanjing 210029, China. E-mail: [mark19900209@163.com, yuansongtao@vip.sina.com, liuqh@njmu.edu.cn].*
- *Q. Chen is with the School of Computer Science and Engineering, Nanjing University of Science and Technology, Nanjing 210094, China. E-mail: chen2qiang@njust.edu.cn.*

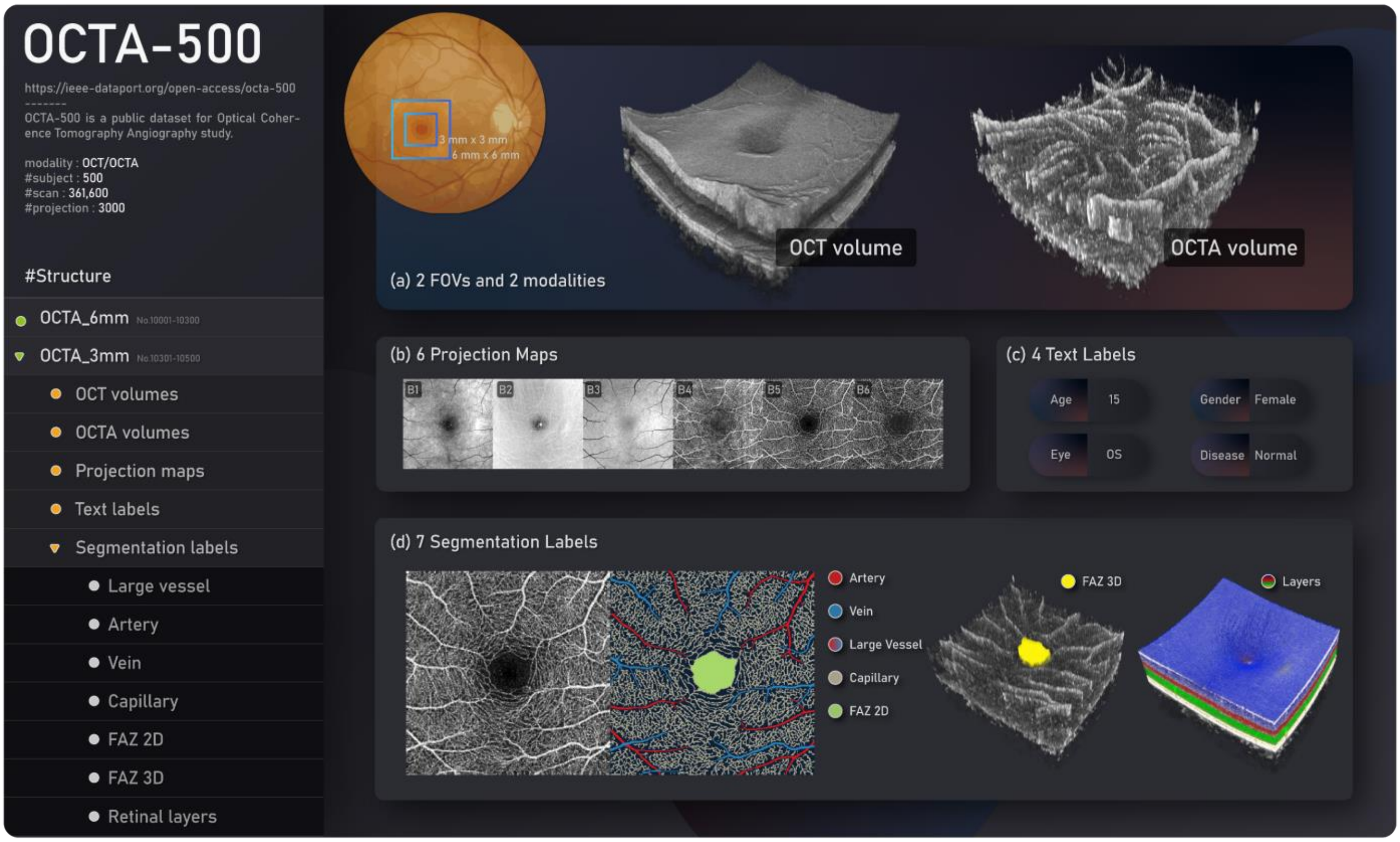

Fig. 1. The structure and contents of the OCTA-500 dataset.

using public datasets (Owen et al., 2009; Staal et al., 2004; Hoover et al., 2002). Due to the late start of OCTA, available OCTA datasets are few and small. Image sources from OCTA imaging remain precious and scarce. The limited size of currently available datasets is a major challenging factor (Yao et al., 2020). We summarize four available OCTA datasets (Giarratano et al., 2020; Ma et al., 2020; Díaz et al., 2019; Agarwal et al., 2020) in Table 1, which has the following limitations: (1) The number of images and the disease diversity are limited. (2) The datasets are single-modality, and the datasets only provide OCTA projection maps. 3D volumes of OCT and OCTA are rare and urgently needed by researchers. (3) The datasets are single-task datasets, providing only one type of label. The dedicated study for a single task is commendable. Nevertheless, it is necessary to integrate the studies. (4) The latest foci, such as 3D FAZ segmentation and artery-vein segmentation, have no available public datasets.

To address these limitations and guide OCTA research forward, we introduce a new well-organized OCTA dataset named OCTA-500. Fig. 1 shows the content and organization of OCTA-500. The images for this dataset were collected by an OCTA device from 500 subjects. The dataset provides rich images and annotations including two fields of view (FOVs, 6 mm/3 mm), two modalities (OCT/OCTA volumes, 361,600 scans), six types of projections, four types of text labels (age/gender/eye/disease) and seven types of segmentation labels (large vessel/capillary/artery/vein/2D FAZ/3D FAZ/retinal layers). This dataset is one of the few that contain OCT/OCTA volumes. The size of OCTA-500 (>80 GB) far exceeds that of other OCTA datasets (<0.2 GB in Giarratano et al., 2020; Ma et al., 2020; Díaz et al., 2019; Agarwal et al., 2020). The segmentation labels we annotated can be used to explore the possible performance improvement and space optimization brought by multi-task collaboration. Among these labels, artery labels, vein labels and 3D FAZ labels in OCTA images are given publicly for the first time. Without any exaggeration, OCTA-500 is currently the largest and most comprehensive OCTA dataset.

Based on the segmentation annotations in OCTA-500, we propose a novel multi-object segmentation task, called CAVF, which integrates capillary segmentation, artery segmentation, vein segmentation, and FAZ segmentation under a unified framework. The proposed CAVF will bring convenience in the computation of quantitative indicators and the evaluation of model performance. Focusing on the CAVF task, several state-of-the-art 2D segmentation networks were selected as baselines. In addition to 2D segmentation networks, we also considered the 3D-to-2D segmentation method image projection network (IPN) (Li et al., 2020a). To achieve competitive segmentation performance, we further optimized the IPN to IPN-V2, to serve as one of the 3D-to-2D segmentation baselines for the CAVF task.

By means of a series of experiments on OCTA-500, we address several questions: How much do deep learning methods improve with an increased amount of training data? How do different modality inputs affect segmentation quality? Which baselines perform well on the CAVF task? How well do segmentation models perform in different diseases?

**Table 1**
Summary of public OCTA datasets.

| | Giarratano et al. (2020) | ROSE (Ma et al., 2020) | OCTAGON (Díaz et al., 2019) | FAZID (Agarwal et al., 2020) | **OCTA-500** |
|---|---|---|---|---|---|
| #Modalities | 1 | 1 | 1 | 1 | **2** |
| #Subjects | 11 | 151 | 213 | 304 | **500** |
| #Diseases | 1 | - | 2 | 3 | **>12** |
| #Projections | 1 | 3 | 4 | 1 | **6** |
| #Texts | 0 | 0 | **4** | **4** | **4** |
| Volumes | - | - | - | - | **√** |
| Segmentations | Capillary | Capillary | 2D FAZ | 2D FAZ | **Large vessel; capillary; artery; vein; 2D/3D FAZ; layers** |
| FOV ($mm^2$) | 3 × 3 | 3 × 3 | **6 × 6**<br>**3 × 3** | 6 × 6 | **6 × 6**<br>**3 × 3** |
| Resolution | (91, 91) | **(304, 304)**<br>**(512, 512)** | (320, 320) | (420, 420) | **(640, 400, 400)**<br>**(640, 304, 304)** |

*#Diseases: Statistics of disease types include healthy cases.*

## 2 Related Work

### 2.1 OCTA Dataset

Over the past two decades, the continuous emergence of color fundus image datasets, such as CHASE (Owen et al., 2009), DRIVE (Staal et al., 2004), and STARE (Hoover et al., 2002), has stimulated enthusiasm for retinal research. OCTA is a relatively novel imaging modality. Due to the late start, only a small number of OCTA datasets are publicly available thus far. See Table 1 for an overview of the current public OCTA datasets. Giarratano (Giarratano et al., 2020) and ROSE (Ma et al., 2020) were used for vessel segmentation, while OCTAGON (Díaz et al., 2019) and FAZID (Agarwal et al., 2020) were used for FAZ segmentation. These existing datasets focus on a single specific task, and no multi-task OCTA dataset has yet been published.

In terms of data diversity, we focus on imaging modality (OCT/OCTA), data structure (3D volumes/projection maps), and field of view (FOV, 6 mm/3 mm). Giarratano, ROSE, OCTAGON, and FAZID are all single-modality datasets. They contain only OCTA modality, to be precise, only projection maps of OCTA. ROSE-1 and OCTAGON include superficial and deep projection maps, while Giarratano and FAZID only include superficial projection maps. These projection maps are generated from 3D OCTA volumes, and OCTA volumes are derived from OCT volumes (Jia et al., 2012). Multiple modalities and 3D volume data are required in many OCTA studies (Li et al., 2020a; Xu et al., 2022; Vogl et al., 2017; Jia et al., 2014; Zhang et al., 2016; Zhang et al., 2020; Lee et al., 2019; Yang et al., 2020; Zhang et al., 2017). Regrettably, none of these studies contain OCT modality and 3D volume data. In FOV, OCTAGON includes both 6 mm × 6 mm and 3 mm × 3 mm FOVs, while others contain only 3 mm × 3 mm FOV.

Disease diversity is another important aspect. Disease samples can better reflect the generalization performance of different methods, so datasets with disease diversity can be more widely used. FAZID had the largest number of subjects (= 304) in the existing dataset, but it contains only two diseases, diabetic retinopathy (DR) and myopia. ROSE-2 contains various macular diseases, but the disease label for each image is not explicitly given. OCTAGON contains only DR data. Giarratano considers only normal cases and does not include subjects with any diseases. Some common retinal diseases, such as age-related macular degeneration (AMD), choroidal neovascularization (CNV), central serous chorioretinopathy (CSC), retinal vein occlusion (RVO), etc., are not included in the existing OCTA datasets.

### 2.2 Vessel Segmentation

We have also witnessed impressive achievements in retinal blood vessel segmentation for color fundus images (Moccia et al., 2018; Mookiah et al., 2021; Fraz et al., 2012). The OCTA imaging modality allows visualization of the microvasculature (Kashani et al., 2017) and has been considered a powerful tool to observe retinal vessels (Spaide et al., 2018). Vessel segmentation in OCTA presents a series of challenges due to noise, poor contrast, low resolution, and high vessel complexity (Mou et al., 2021; Ma et al., 2020). Nonetheless, benefiting from the inheritance of previous vessel segmentation methods on color fundus images and the development of deep learning techniques, vessel segmentation in OCTA images has achieved rapid development in recent years.

Vessel segmentation methods in OCTA images can be divided into threshold-based, filtering-based, active contour model-based, and deep learning-based methods. Threshold-based methods commonly used in OCTA images are summarized by Terheyden et al. (2019) and Mehta et al. (2020). The vasculature has a higher signal intensity than background tissues in OCTA images, so using a threshold can simply separate blood vessels and backgrounds. However, as mentioned in (Li et al., 2017), an obvious weakness of threshold-based methods is their intolerance to background noise. Filtering methods have thus been considered for noise suppression and vessel enhancement. In studies (Chu et al., 2016; Xu et al., 2019; Aharony et al., 2019), a Frangi filter (Frangi et al., 1998) was applied to enhance blood vessels, and then different threshold-based methods were used to obtain vessel binary images.

Li et al. (2017) adopted top-hat enhancement and optimally oriented flux (OOF) for small vessel detection. Threshold-based and filter-based methods can roughly segment vessels for estimating vessel density and vessel skeleton density (Terheyden et al., 2019). However, their segmentation usually performs poorly in the presence of artifacts, motion blur, weak contrast, low resolution, and disease (Mou et al., 2019). In addition, specific structures such as capillaries, arteries, and veins are also difficult to discriminatively segment due to their lack of sufficient recognition capabilities.

Recently, the use of deep learning frameworks has stimulated the rapid development of vessel segmentation algorithms. U-Net (Ronneberger et al., 2015) is a typical encoder-decoder structure with multiscale feature representation and is used by several works (Giarratano et al., 2020; Mou et al., 2021; Pissas et al., 2020; Lo et al., 2020) for vessel segmentation in OCTA images. The CS-Net proposed by Mou et al. (2021) supplements a spatial attention module and a channel attention module in the U-Net structure and was used to segment the blood vessel skeleton in OCTA images. Giarratano et al. (2020) compared 8 different thresholding, filtering, and deep learning methods on a small dataset with 55 region of interest (ROI) slices. They showed that U-Net and CS-Net achieve the best performance (Dice = 0.89) and OOF is the best filter (Dice= 0.86) on their capillary segmentation task. Ma et al. (2020) more recently presented OCTA-Net with two split-based stages to detect thick and thin vessels separately. The above methods focus on segmenting 2D projection maps, whose generation relies on accurate layer segmentation. Our previous work (Li et al., 2020a) introduced an image projection network (IPN), which provides a 3D-to-2D segmentation approach to segment 2D large vessels from 3D OCTA volumes. This method does not require the use of layer segmentation to generate a 2D projection map. Based on the IPN structure, the recent works PAENet (Wu et al., 2021) and PRSNet (Li et al., 2022c) supplement the quadruple attention module and dual-way projection learning module respectively.

In addition to segmenting vessels with different calibers, Alam et al. (2018, 2019) also showed that artery-vein segmentation in OCTA images is feasible and a potential research direction. They used a fully convolutional network named AV-Net (Alam et al., 2020) to segment the arteries and veins in OCTA images. However, due to the lack of public annotations, artery-vein segmentation in OCTA images has not been widely developed. To fill this vacancy and stimulate its development, the artery-vein segmentation annotations are included in the OCTA-500 dataset.

### 2.3 FAZ Segmentation

The foveal avascular zone (FAZ) is a nonperfused region of the fovea, which is surrounded by interconnected retinal vessels. The FAZ also approximates the region of highest cone photoreceptor density and oxygen consumption. Over the last decade, the FAZ has often been analyzed under different aspects, mainly by fluorescence angiography (FA) (Conrath et al., 2005; Zheng et al., 2010). The novel OCTA technology allows a noninvasive examination, visualization and quantitative analysis of the FAZ. Several new findings in OCTA reveal that the size of the FAZ is highly correlated with both visual acuity and disease (Freiberg et al., 2016; Balaratnasingam, 2016). Hence, the FAZ has received increasing attention from clinicians and researchers, and its segmentation has become a focus in OCTA studies. Quantitative evaluation of the FAZ area and perimeter relies on FAZ segmentation. FAZ segmentation is also used to locate the foveal center (Li, 2020b). A more recent study (Lin et al., 2021) also shows that the features in FAZ segmentation can also be used to improve the performance of multi-disease classification.

Current FAZ segmentation methods can be divided into unsupervised methods and supervised methods. The unsupervised methods are designed according to the location characteristics (located at the imaging center), geometric characteristics (unique connected area) and gray characteristics (no vessel) of the FAZ. They often need to set the initial seed or initial contour. For example, Lu et al. (2018) presented a generalized gradient vector flow (GGVF) snake model to evolve the FAZ contour through an initial circular area. Xu et al. (2019) determined initial seeds in distance transform images and then used a graph cut to finish segmentation. These unsupervised methods could have excellent performance on a small dataset with a healthy or single disease but might encounter more difficulties when applied to larger and more complex datasets (Guo et al., 2021). Deep learning techniques have demonstrated an excellent ability to determine the complex structure of high-dimensional data. Many deep learning methods have been reported for FAZ segmentation. The U-Net backboned encoder-decoder structure is currently the most used network architecture in the FAZ segmentation task and has been adopted by many works (Guo et al., 2019b, 2021; Li et al., 2020b; Guo et al., 2018, 2019a; Liang et al., 2021; Jabour et al., 2021) for end-to-end FAZ segmentation on 2D OCTA projection maps (enface OCTA). Mirshahi et al. (2021) used Mask R-CNN for FAZ segmentation on enface OCTA images. Our previous work also performed 3D-to-2D FAZ segmentation by using IPN (Li et al., 2020a). Xu et al. (2021) more recently proposed a new index, FAZ volume. To calculate the FAZ volume, they developed a 3D U-Net structure to segment the 3D FAZ (Xu et al., 2022). It is worth mentioning that both 2D and 3D FAZ annotations are included in OCTA-500. More related works on vessel segmentation and FAZ segmentation are summarized in Tables S1 and S2 in Supplementary Material.

### 2.4 Main Contributions

Our contributions are threefold:

- We introduce the OCTA-500 dataset, which contains OCTA imaging under two FOVs from 500 subjects. The OCTA-500 dataset provides rich images and annotations including 2 modalities (OCT/OCTA volumes), 6 projections, 4 text labels (age/gender/eye/disease), and 7 segmentation labels (large vessel/capillary/artery/vein/2D FAZ/3D FAZ/retinal layers). It is currently the most comprehensive OCTA dataset. See Section 3.
- We propose a CAVF task that integrates artery

segmentation, vein segmentation, capillary segmentation, and FAZ segmentation under a unified framework. Based on the CAVF task, we optimize the 3D-to-2D network IPN to IPN-V2 to serve as one of the competitive baselines. See Section 4.

- We provide insights into several dataset characteristics: the training set size, the model input (OCT/OCTA, 3D volume/2D projection), the baselines (2D-to-2D baselines/3D-to-2D baselines), and the diseases. See Section 5.

## 3 OCTA-500 Dataset

In this section, we first introduce the data collection process and the two subsets of the OCTA-500 (Section 3.1); we then introduce the contents of the OCTA-500 including OCT/OCTA volumes (Section 3.2), projection maps (Section 3.3), text labels (Section 3.4) and segmentation labels (Section 3.5).

### 3.1 Data Collection

OCTA-500 contains a total of 500 subjects divided into two subsets according to the FOV type: OCTA_6mm and OCTA_3mm. OCTA_6mm includes 300 subjects (NO. 10001 - NO. 10300) who underwent macular OCT/OCTA imaging with a FOV of 6 mm × 6 mm. OCTA_3mm includes 200 subjects (NO. 10301 - NO. 10500) who underwent macular OCT/OCTA imaging with an FOV of 3 mm × 3 mm. The OCT/OCTA images are from the same device, which is a commercial 70 kHz spectral-domain OCT system with a center wavelength of 840 nm (RTVue-XR, Optovue, CA). All subjects were imaged at Jiangsu Province Hospital from March 2018 to July 2020. To ensure patient independence, only one eye of each subject was included. The registration information of all subjects was complete, and the type of diagnosed disease was given by ophthalmologists. OCTA_6mm and OCTA_3mm came from two independent recruitments and therefore differed in disease diversity. The subjects of OCTA_6mm are mainly from a population with common retinal diseases, and the normal population is a minority. The subjects of OCTA_3mm are mainly from the normal population, and secondarily from the diseased population. More details about the disease distribution can be found in Section 3.4.

### 3.2 OCT and OCTA Volumes

The OCTA-500 dataset contains volume data of two modalities, OCT and OCTA volumes, as shown in Fig. 2. OCT volumes and OCTA volumes provide structural information and blood flow information of the retina, respectively. The OCTA volume is generated from OCT volumes by the split-spectrum amplitude-decorrelation angiography (SSADA) algorithm (Jia et al., 2012). Due to the use of band-pass filtering and split-spectrum methods in the SSADA algorithm, the vertical resolution of the OCTA volume is 1/4 of the OCT volume. We used bilinear interpolation to resize it, and then the OCT and OCTA volumes were registered. The FOV and resolution of the two subsets are different. The imaging range of OCTA_6mm is 6m m × 6 mm × 2 mm centered on the fovea, and its volume size is

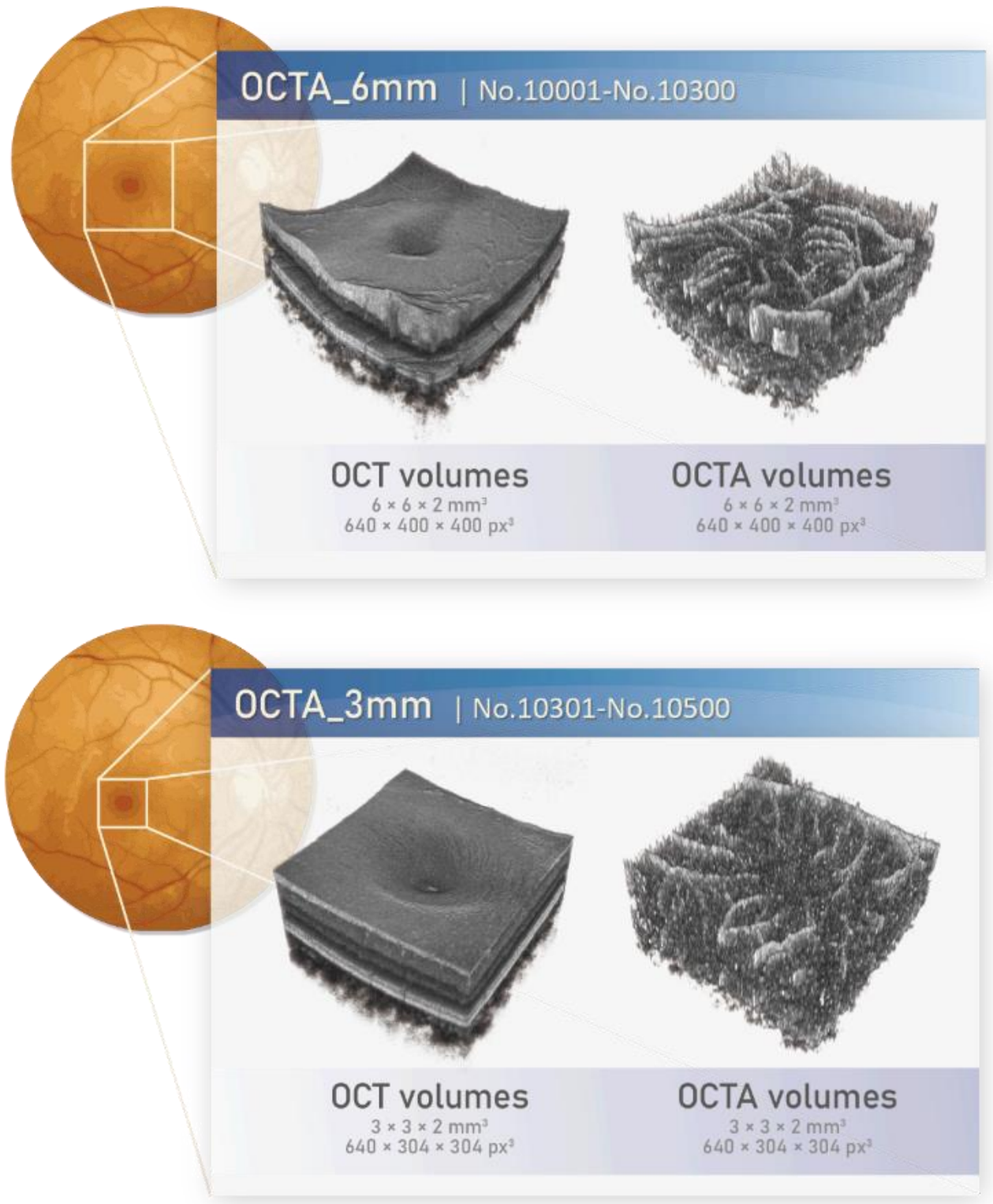

Fig. 2. 3D visualization of the OCT volumes and OCTA volumes in two fields of view.

400 px × 400 px × 640 px. The imaging range of OCTA_3mm is 3 mm × 3 mm × 2 mm centered on the fovea, and the volume size is 304 px × 304 px × 640 px. For ease of reading by researchers, we present these 3D volumes in the form of 2D scans (B-scans), for a total of 361,600 scans.

### 3.3 Projection Maps

The provision of 3D OCT/OCTA volumes allows us to separately project and visualize different retinal layers through layer segmentation. In OCTA-500, we provide six types of projection maps, as shown in Fig. 3. We used the retinal layer position information of the internal limiting membrane (ILM) layer, the outer plexiform layer (OPL), and Bruch's membrane (BM). Two kinds of projection methods are selected for the generation of projection maps: average projection and maximum projection. These projection methods are obtained by averaging or maximizing along the axial direction. The OCT volume usually uses the average projection. Since the value in OCTA volume reflects the intensity of the blood flow signal, to show the shape of blood vessels more clearly, the maximum projection is usually used in the projection maps of the inner retina and outer retina (Hormel et al., 2018). The projection maps we generated are as follows:

(B1) OCT full projection, the average value of 3D OCT volume along the axial direction, which shows the global information of the OCT volume. OCT full projection is often used to observe retinal vessels and edema (Vogl et al., 2017).

(B2) OCT average projection between the ILM and OPL. It can show the vessels in the inner retina with high

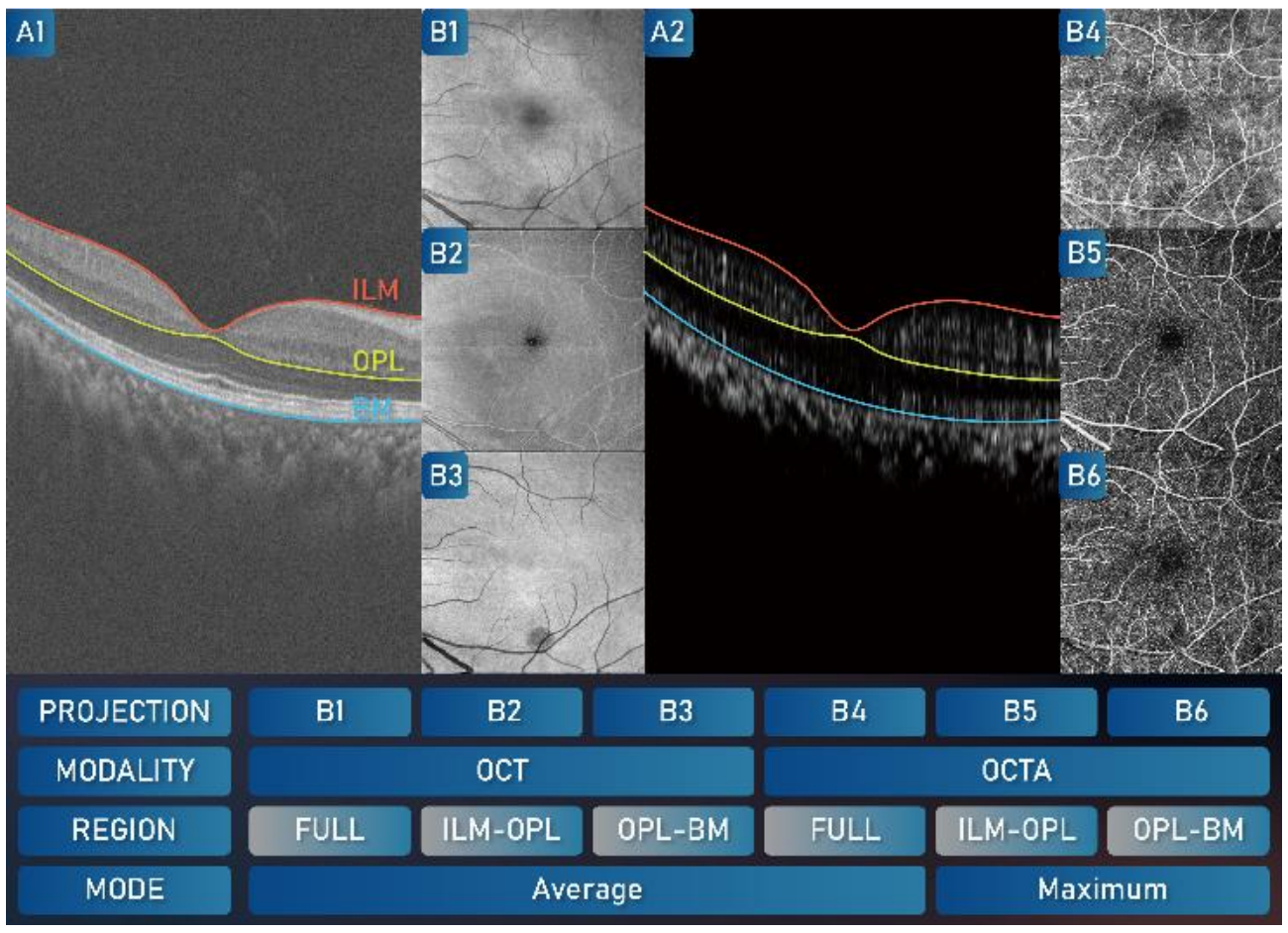

Fig. 3. Generation of the projection maps. (A1-A2) A B-scan of OCT/OCTA and the layer segmentation used. (B1-B3) OCT projection (B4-B6) OCTA projection. The table lists the modality, projection region and projection mode of each projection map.

reflection (Chen et al., 2016).

(B3) OCT average projection between the OPL and BM. It displays the vessel shadows in the outer retina with low reflection. It shows higher vessel contrast than the B2 projection (Chen et al., 2016).

(B4) OCTA full projection. It is the average value of 3D OCTA volume along the axial direction, which is a global view of both the retina and choroid.

(B5) OCTA maximum projection between the ILM and OPL. It is generated by the maximum projection of the inner retina which can clearly show the vascular morphology of the inner retina (Li et al., 2020a) and the shape of the FAZ (Lu et al., 2018).

(B6) OCTA maximum projection between the OPL and BM. It is generated by the maximum projection of the outer retina, which can be used to observe and monitor the morphology of CNV (Jia et al., 2014; Zhang et al., 2016).

Note that the projection maps we have given above are common, but not all. We also provide layer segmentation annotations (see details in Section 3.5.5) so that researchers can generate projection maps according to their needs.

## 3.4 Text Labels

To count the data distribution and disease diversity in OCTA-500, we sorted four text labels from the medical records as follows: (a) gender, (b) eye, (c) age, and (d) disease. Their distributions are shown in Fig. 4. The average ages of the subjects included in the OCTA_6mm and OCTA_3mm are 49.18 ± 17.28 and 33.12 ± 16.17, respectively. OCTA_6mm has richer disease diversity than OCTA_3mm. The proportion of subjects with ophthalmic diseases in OCTA_3mm is 20%, and the diseases included are AMD, DR, and CNV. The proportion of subjects with ophthalmic diseases in OCTA_6mm is 69.7%. The diseases include AMD, DR, CNV, CSC, RVO, and others. ‘Others’ here refers to diseases with a small number (n<8), including retinal detachment (RD), retinal hemorrhage (RH), optic atrophy (OA), epiretinal membrane (ERM), retinitis pigmentosa (RP), central retinal artery occlusion (CRAO), retinoschisis, etc.

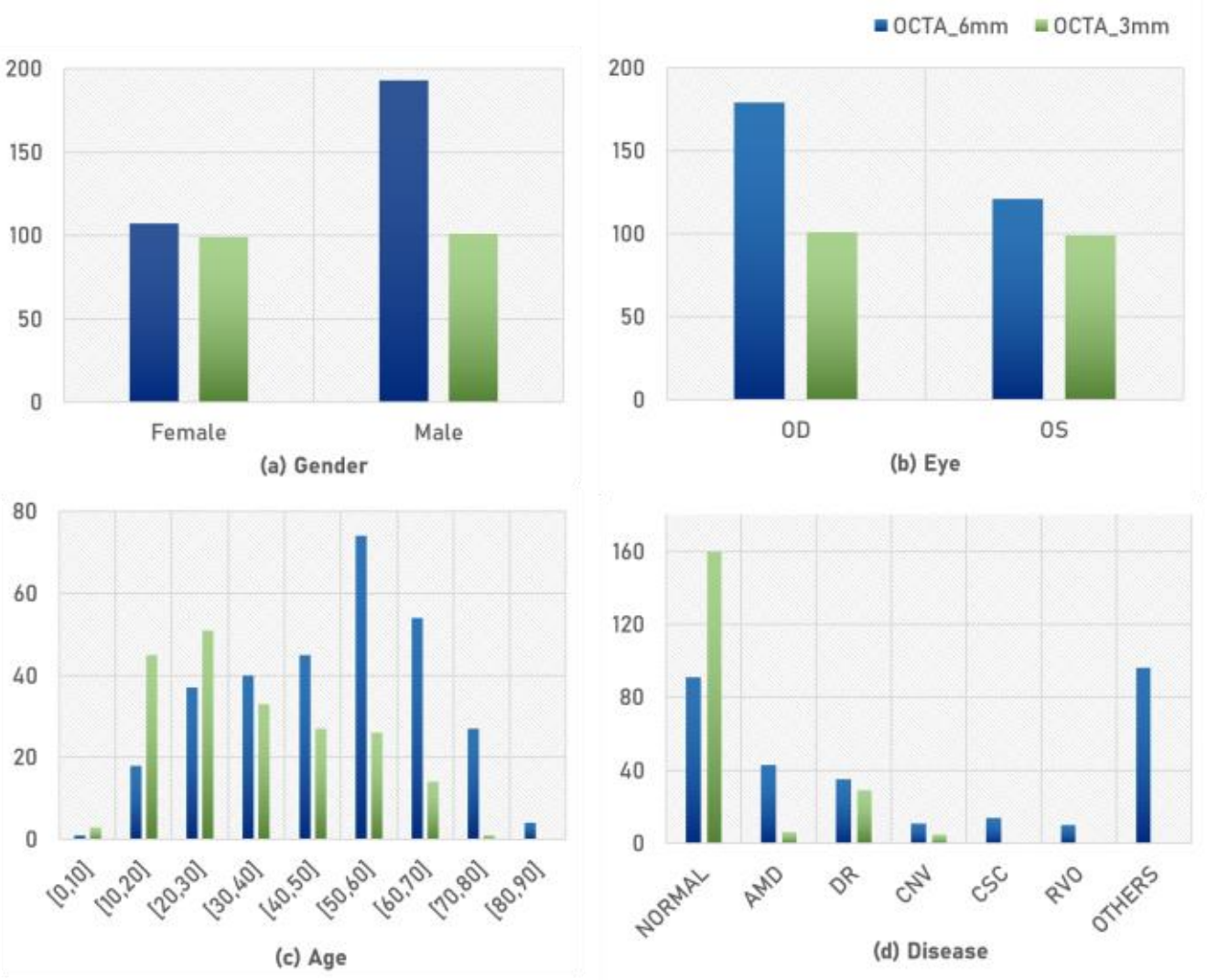

Fig. 4. Statistical histogram of text labels in OCTA-500: (a) gender, (b) eye, (c) age, (d) disease. ‘OD’ means right eye, ‘OS’ means left eye.

## 3.5 Segmentation Labels

In this section, we introduce the labeling criteria and processes of the segmentation labels in OCTA-500, including large vessel, artery, vein, capillary, 2D/3D FAZ, and retinal layers.

### *3.5.1 Large Vessel*

There have been many applications for segmenting large vessels in OCTA images. For example, the vessel density of large vessels is an important indicator in assessing retinal disease (Xu et al., 2019). Large vessels can be used as a mask to remove artifacts, which is a necessary step for segmenting capillaries in OCTA images (Jiang et al., 2018; Xu et al., 2019; Li et al., 2017). In Section 3.5.2, we also show that large vessels can be further differentiated into arteries and veins, motivating further development of the vascular assessment. In this section, we will introduce how we accurately label such important large vessels.

Large vessels are distributed in the inner retina, showing a high-intensity tree-shaped structure in OCTA projection map B5 (Fig. 5a). The signal intensity of large vessels is generally slightly higher than that of capillaries. However, it is still difficult to segment it accurately only by threshold binarization. Fig. 5e shows the result of segmenting large vessels with threshold binarization, which still retains capillaries and noise. To obtain accurate and clean large vessel segmentation labels, we perform manual labeling of large vessels.

We used Adobe Photoshop CC (Adobe Systems, Inc., San Jose, CA, USA) to annotate the large vessel in projection map B5. The labeling process is divided into two steps: coarse-grained annotation and fine-grained correction. In the coarse-grained annotation stage, we draw the large vessels using a red brush (R: 255, G: 0, B: 0, hardness: 100) on a separate layer. The thickness of the brush is larger than the diameter of the blood vessel. In this stage, we do not pursue the precise boundary, but ensure that there are no missing blood vessels. The result of this annotation stage has a thicker vessel diameter, as shown in Fig. 5b. We then performed fine-grained correction delicately delineate the vessel boundaries. In this time, we used the 'Screen' mode, in which the over-segmented part will appear red with higher saturation (Fig. 5c, yellow arrow). We finely delineated the boundary so that the shape of the blood vessels was consistent with the gray distribution in the projection map, and at the same time, the smoothness and continuity of the blood vessels were also ensured (Fig. 5d). Return to "normal" mode, adjust the color, and obtain the final large vessel label (Fig. 5f).

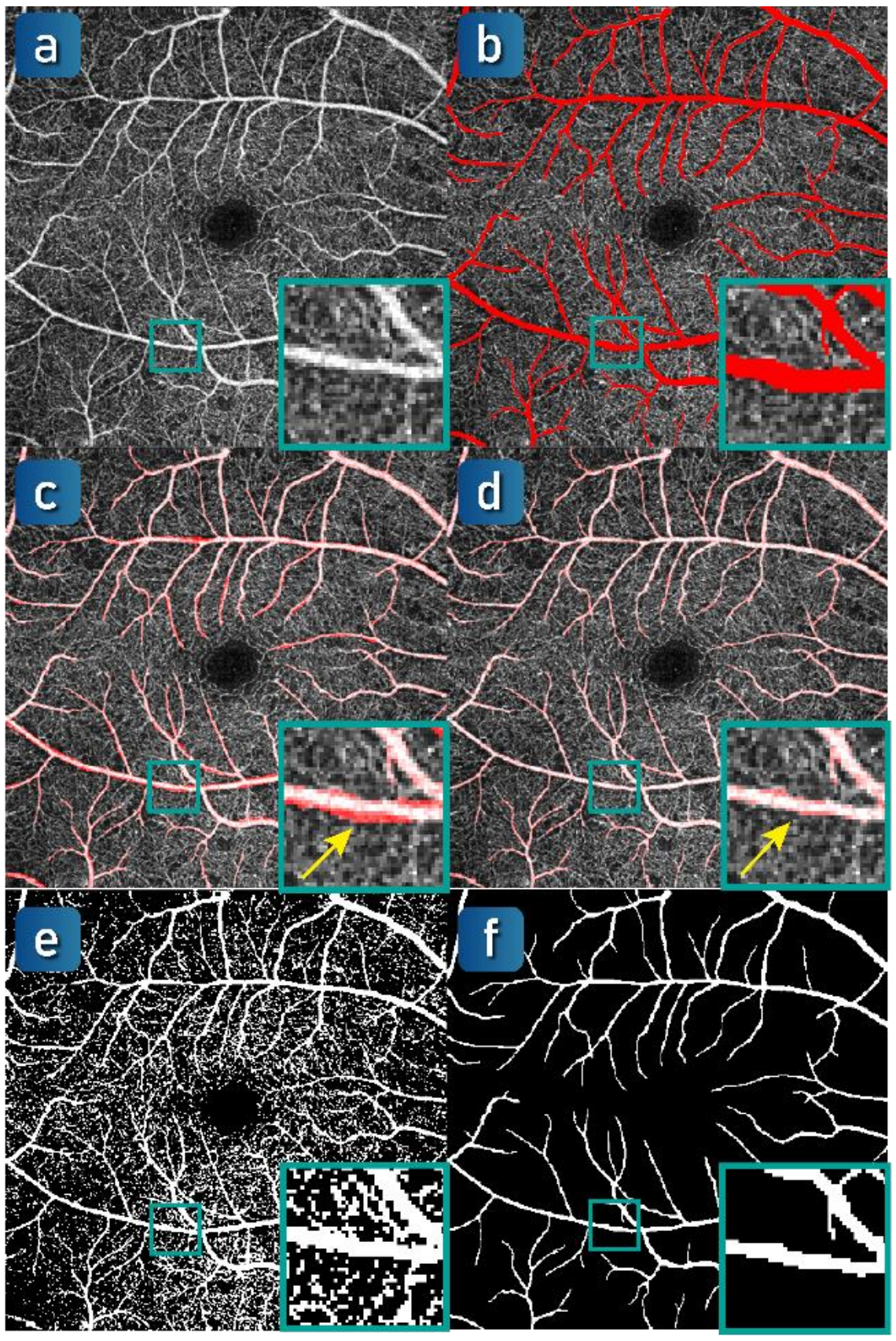

Fig. 5. Annotation of the large vessels in OCTA-500. (a) OCTA projection map (B5). (b) Coarse-grained manual annotation. (c) Visualization of coarse-grained annotation in 'Screen' mode. (d) Fine-grained manual corrections in 'Screen' mode. (e) Threshold result. (f) The final label of large vessels.

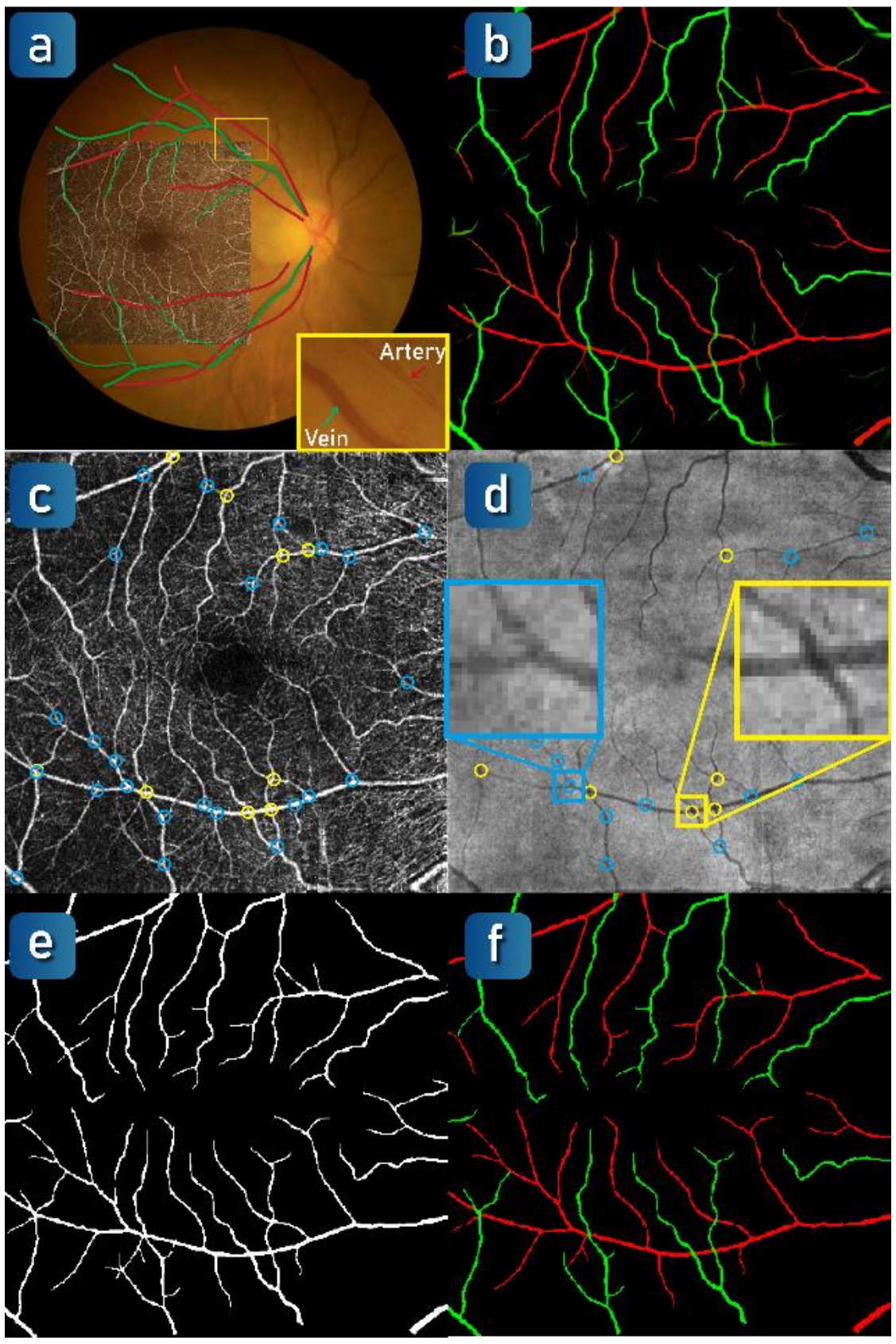

Fig. 6. Annotation of the arteries and veins in OCTA-500. (a) Color fundus image. (b) The mean results of models trained on an extra dataset. (c) OCTA projection map (B5). (d) OCT projection map (B3). Blue circles represent branch points and yellow circles represent crossover points. (e) The large vessel label. (f) The final artery-vein label. Red represents arteries. Green represents veins.

### *3.5.2 Artery and Vein*

Large vessels can be further divided into arteries and veins. Differential artery-vein analysis can provide valuable information for evaluating ophthalmic diseases and improving the performance of disease classification (Alam et al., 2019a). At present, very few studies segment arteries and veins in OCTA images. Alam et al. (2018, 2019a, 2019b) recently demonstrated the potential of differentiating arteries and veins in OCTA. One of their key proposals is to use color fundus images as a guide to distinguish between arteries and veins in OCTA images. Indeed, even for human experts, it is difficult to manually annotate arteries and veins using only OCTA images. To improve the efficiency and accuracy of artery-vein annotation, we refer to multiple imaging modalities, including color fundus images, OCT, and OCTA.

We divide the manually annotated large vessels into arteries and veins. To this end, we first determined the arterial and venous categories of the main vessels on the color fundus images (Fig. 6a). Annotation guidelines can be found in (Kondermann et al., 2007). Compared with OCT/OCTA, color fundus images present a wider field of view and have color information, making it easier to distinguish between arteries and veins. However, using only color fundus images cannot label the arteriovenous properties of all large vessels because of the limited ability of color fundus images to image vessel branches. We then determined the arteriovenous properties of each vessel branch by identifying vessel crossover points and branch points in OCTA projection maps B5 (Fig. 6c) and B3 (Fig. 6d). The guidelines of the crossover points and branch points are as follows: The branch point is 3-branched and the crossover point is 4-branched; The crossover points usually have a darker color (Fig. 6c, yellow box), and the branch points are more consistent in color (Fig. 6c, blue box); Vessel properties at branch points are consistent; Vessels with the same property generally do not intersect, which means that the two cross vessels are generally contain one artery and one vein.

The properties of the main blood vessels were determined by the color fundus images, and the crossover and branch points were determined by the OCT/OCTA projection maps. Combining the two, the large vessels annotated

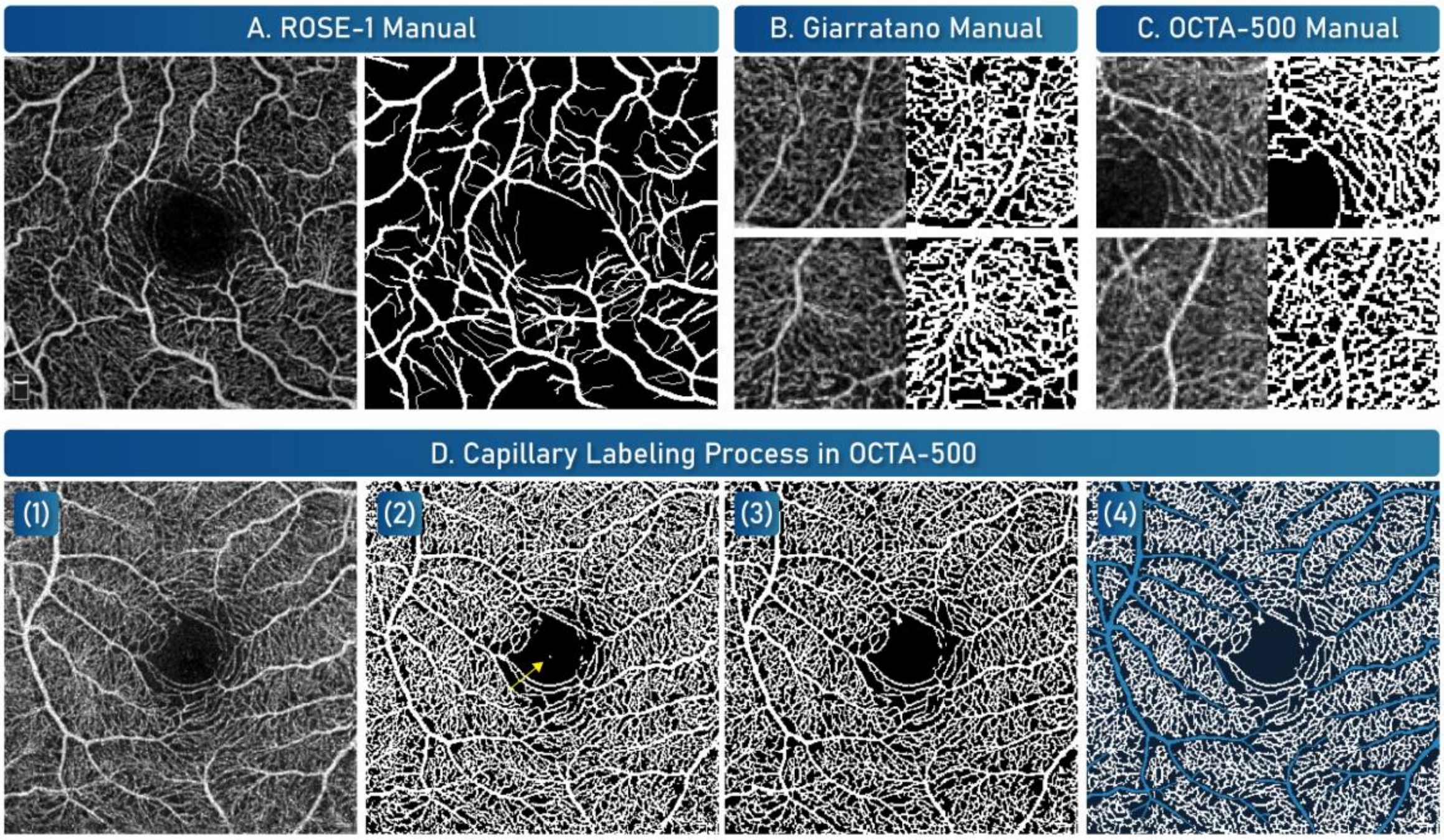

Fig. 7. Annotation of the capillary in OCTA-500. (A) An example of manual annotation from ROSE-1. (B) Examples of manual annotation in Giarratano. (C) Examples of manual labeling of OCTA-500 slices. (D) Capillary labeling process in OCTA-500: (1) OCTA projection map (B5). (2) The preliminary segmentation result using IMN. (3) The result of topology optimization and denoising using LAL. (4) The final label. Blue represents the large vessels. White represents the capillaries.

in Section 3.5.1 (Fig. 6e) can be further marked as arteries and veins.

Numerous subjects in the OCTA-500 were unable to distinguish the arteriovenous properties of the main vessels due to the lack of corresponding color fundus images. To remedy this lack, we trained the deep models on an extra private dataset, which contains 100 subjects with color fundus images and OCT/OCTA images. Arteries and veins are labeled following the above guidelines. We trained three models U-Net (Ronneberger et al., 2015), U-Net++ (Zhou et al., 2018), Attention U-Net (Oktay et al., 2018), and the mean result is shown in Fig. 6b. This mean result has been very close to the real situation, but to ensure the reliability of the labeling, this result is used only to discriminate the properties of the main vessels. Nevertheless, the branch vessels are determined by the crossover points and branch points. In this way, we labeled the vast majority of large vessel labels as artery labels and vein labels (Fig. 6f), but some vessels with relatively small calibers that could not be identified as arteries and veins were excluded.

### 3.5.3 Capillary

Capillaries appear as a dense mesh structure in OCTA images. Manual annotating capillaries in OCTA images is extremely time-consuming, and limited image resolution and the presence of noise make it nearly impossible to manually label all capillaries in the entire image. For this reason, currently supervised vessel segmentation methods focus mainly on large vessels with limited capillary annotation. An example of ROSE-1 (Ma et al., 2020) is given in Fig. 7A. We can see that numerous capillaries are still under-labeled, and their labeling of capillaries is skeleton-level. To obtain a more complete capillary label at the pixel-level, we followed the work of Giarratano et al. (2020). They cropped 55 ROI slices with 91 × 91 pixels on OCTA projection maps and performed detailed large vessel and capillary annotation (Fig. 7B). Through training, testing, and stitching, complete full-vessel segmentation can be obtained.

In this work, we randomly cropped 100 slices with 76 × 76 pixels from the projection map B5 in OCTA-500. The capillaries in these slices were then manually annotated (Fig. 7C). These annotated slices and the slices in the dataset (Giarratano et al., 2020) were both used as training data. We trained the segmentation model using our previously proposed image magnification network (IMN) (Li et al., 2022a). This network structure can well preserve image details, which is dedicated to capillary segmentation. Using the trained IMN model to test the whole projection map B5 (Fig. 7D-(1)) through a sliding window, we can obtain the full-vessel labels (Fig. 7D-(2)). Nevertheless, this result has several deficiencies. On the one hand, there is still noticeable image noise in the results. For example, the pixel indicated by the yellow arrow in the FAZ is identified as a vessel, which may be related to the lack of global information caused by training on the slices. On the other hand, the topological clarity of capillaries still needs to be improved.

To further remove noise and improve the topological clarity of capillaries, we performed additional optimizations on the IMN segmentation results. First, we remove the noise in the FAZ region using the FAZ mask provided in Section 3.5.4. Then, we enhance the topology of blood vessels using our recently proposed label adversarial learning (LAL) (Li et al., 2022b), a skeleton-level to pixel-level vessel segmentation method that can realize the adjustment of the blood vessel diameter and has a certain denoising performance. More details can be seen in the paper (Li et al., 2022b). To prevent model bias, so that labels can be used for performance evaluation of different methods, the backbone of LAL we use is a secret third-party network. The optimized result is shown in Fig. 7D-(3). We can further remove the large vessels (introduced in Section 3.5.1) to obtain the final capillary labels, as shown in Fig. 7D-(4).

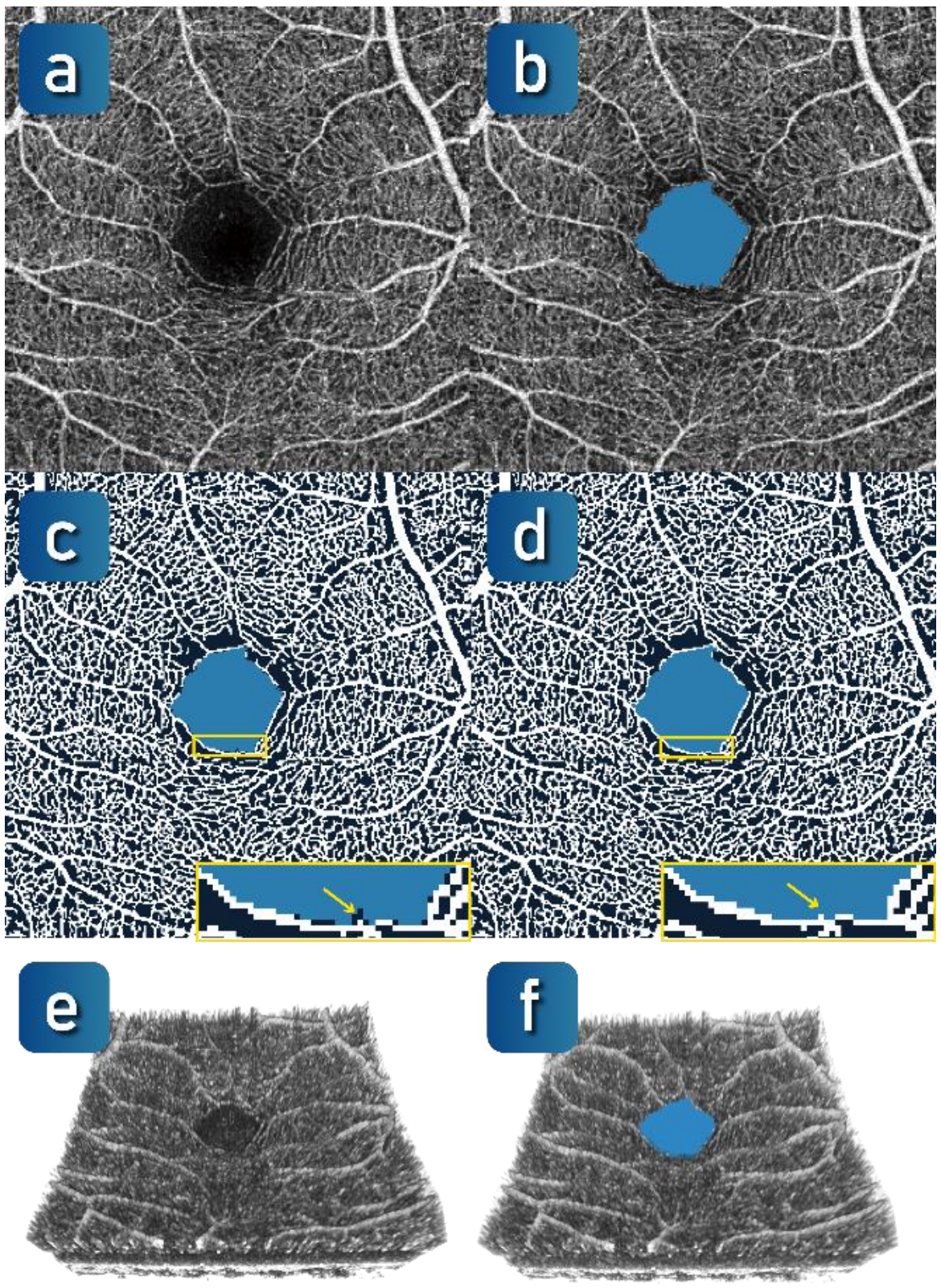

Fig. 8. Annotation of the FAZ in OCTA-500. (a) OCTA projection map (B5). (b) Preliminary manual annotation of the FAZ. (c) Visualization of the FAZ label with the capillary label. (d) The optimized FAZ label. (e) Visualization of the OCTA volume. (f) OCTA volume with 3D FAZ label.

### 3.5.4 Foveal Avascular Zone

In the foveal region, the superficial and deep vascular plexuses form a special capillary-free region, the FAZ, by forming a ring of interconnecting capillaries at the margin of the fovea (Samara et al., 2015). According to this definition, the FAZ should be a 3D region, but at present, most quantification methods for the FAZ are mainly to segment the 2D FAZ on the projection view and calculate 2D indicators, such as the FAZ area. The FAZ area has been found to be associated with visual acuity and disease (Freiberg et al., 2016; Balaratnasingam et al., 2016). Our recent work proposes the FAZ volume and gives a 3D definition of the FAZ (Xu et al., 2021), and it shows a greater sensitivity for vascular alteration. In the OCTA-500 dataset, both 2D and 3D FAZ labels are included.

We annotated the 2D FAZ in the OCTA projection map B5 using Adobe Photoshop CC. The guidelines are as follows: Located in the fovea, generally in the image center; The non-perfused area surrounded by interconnected capillaries; Separate largest closed loop. The quick selection tool is used to improve labeling efficiency. All annotation results were reviewed by multiple experts. Fig. 8b shows an FAZ annotation result. These FAZ labels were published in an earlier version of OCTA-500. In the latest version, we have optimized and updated the FAZ labels. Considering the recently completed capillary labels in Section 3.5.3, we found that the FAZ border did not fit perfectly with the capillary plexus, as indicated by the yellow arrow in Fig. 8c. Thus, we perform a pixel-wise optimization of the FAZ boundaries based on the capillary label, as shown in Fig. 8d. This optimization considers the correlation of the FAZ boundary with the capillary plexus to make the boundary of the FAZ label more accurate. More 2D FAZ annotation examples can be seen in Fig. 10.

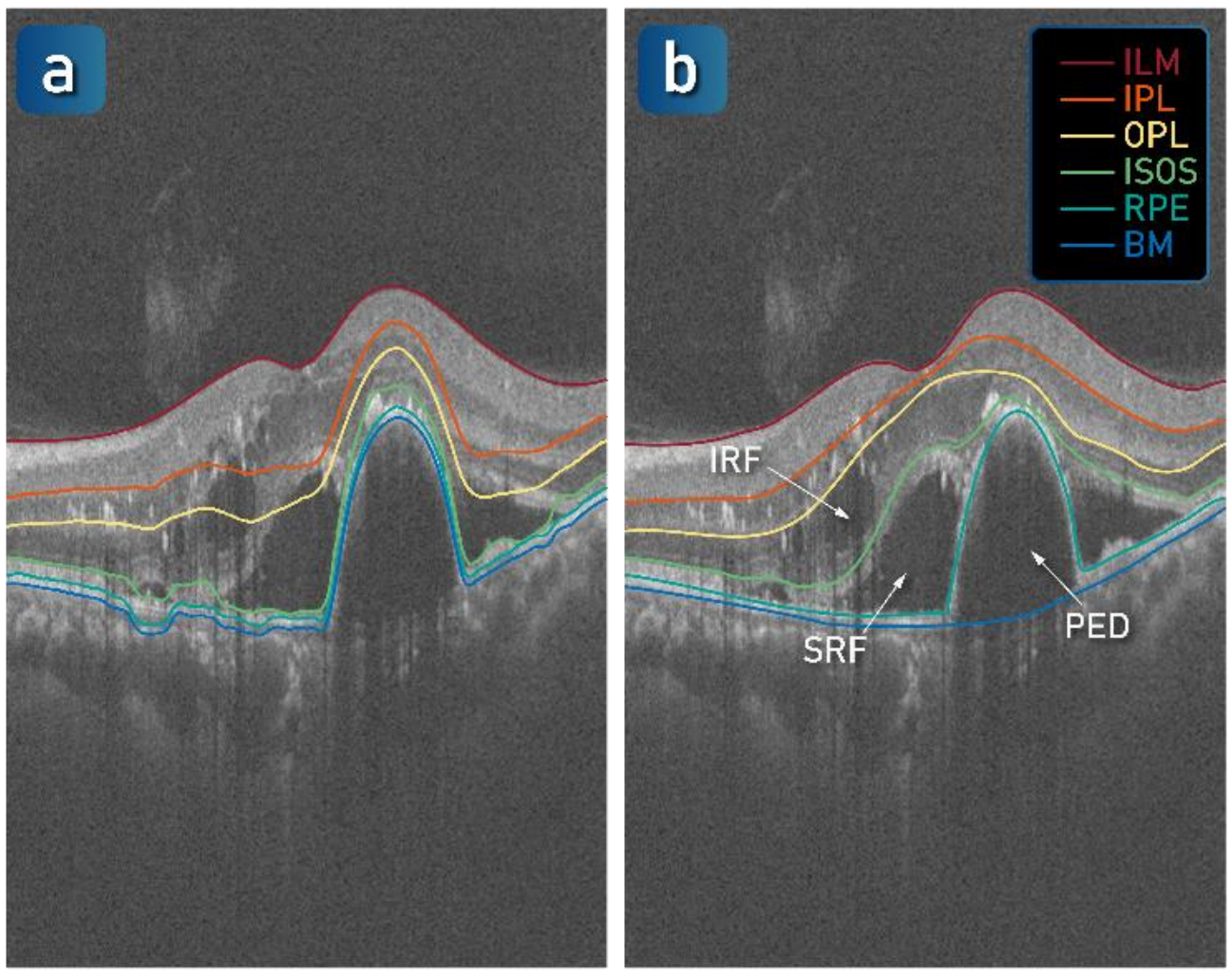

Fig. 9. Annotation of the retinal layers in OCTA-500. (a) An example of layer segmentation from an AMD patient using the Iowa software. (b) The layer labels after manual correction.

The labeling of the 3D FAZ was performed in the OCTA volume (Fig. 8e). The area is limited between the ILM layer and the OPL layer. The layer segmentation used is described in Section 3.5.5. Axial slices were extracted for labeling, and the guidelines were consistent with that of 2D FAZ. More processing details can also be found in the paper (Xu et al., 2021), which used part of the OCTA-500 data as the research object. The 3D FAZ annotation results are shown in Fig. 8f.

### 3.5.5 Retinal Layers

Layer segmentation is an important tool for analyzing the structure and function of different layers in OCT/OCTA images. Retinal thickness analysis (Geobel et al., 2002), vessel density statistics between different layers (Lavia et al., 2019), and generation of projection maps as described in Section 3.3 all require layer segmentation. Numerous automatic retinal layer segmentation algorithms have been reported in recent years (Li et al., 2006; Garvin et al, 2009; Antony et al., 2011; Novosel et al., 2017; Xiang et al., 2018; Zhang et al., 2020; Bogunovic et al., 2019; Yang et al., 2022). Diseases can alter or destroy the retinal layers, so that the retinal layers have complex and diverse shapes under different diseases (Fig. 10), which further leads to the design of retinal layer segmentation algorithms under disease diversity being still a challenging task. To motivate the layer segmentation task to move forward and to facilitate readers to better use OCTA-500, we release the layer segmentation labels.

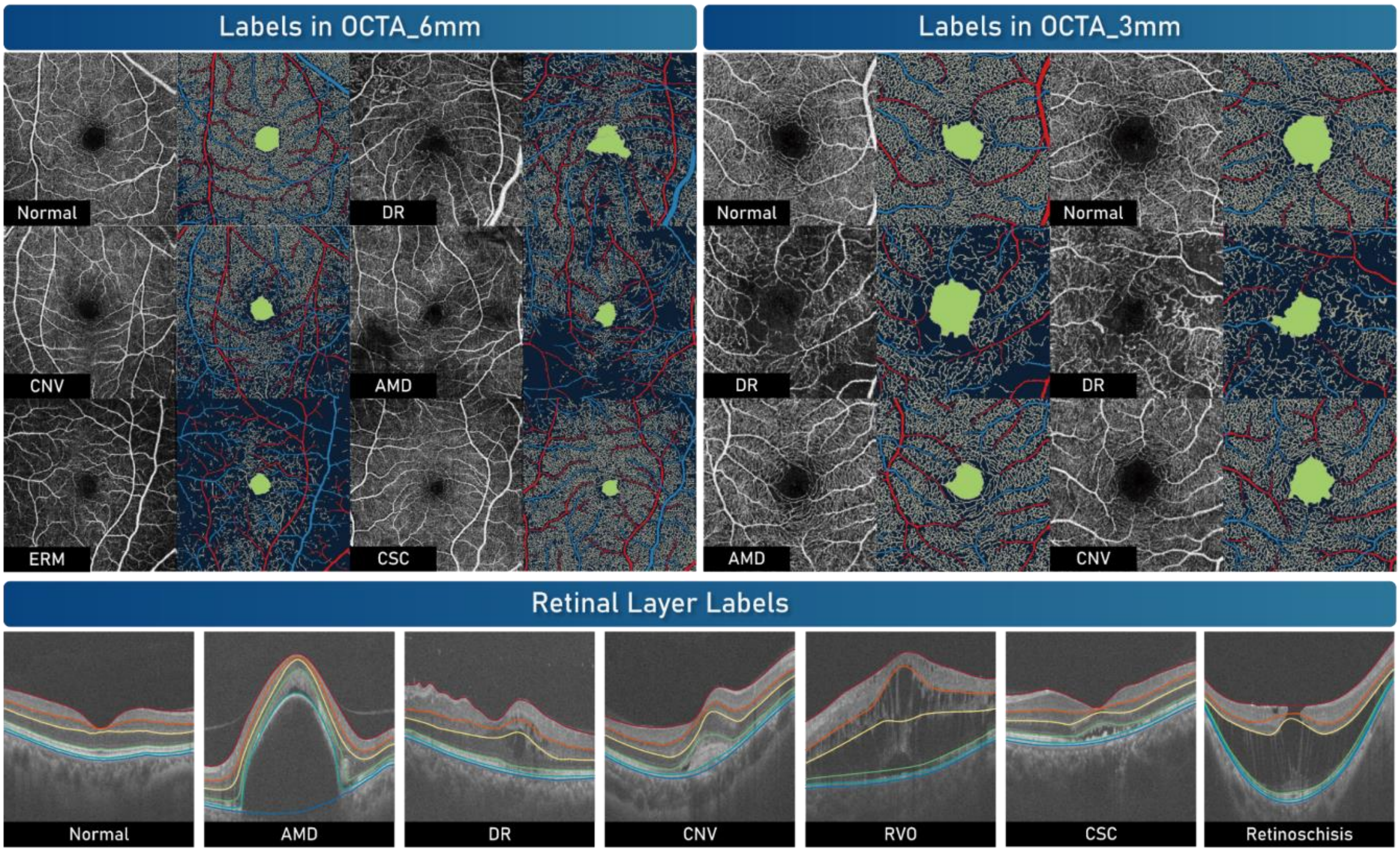

Fig. 10. Diversity of the segmentation labels in the OCTA-500 dataset.

We label 6 retinal layers, which are internal limiting membrane (ILM), inner plexiform layer (IPL), outer plexiform layer (OPL), inner segment/outer segment (ISOS), retinal pigment epithelium (RPE), and Bruch's membrane (BM). We first obtained preliminary results for these 6 layers using Iowa software (OCTExplorer 3.8) (Li et al., 2006; Garvin et al., 2009; Antony et al., 2011). These results are usually inaccurate in disease cases, so we further corrected the results. Fig. 9(a) shows a case of layer segmentation errors under AMD disease. Three types of fluid, intraretinal fluid (IRF), subretinal fluid (SRF), and pigment epithelial detachment (PED), affect the layer structure in this case. We modified the layers of this case referring to the definition in (Bogunovic et al., 2019): SRF is between the neurosensory retina and the underlying RPE that nourishes photoreceptors. PED represents detachment of the RPE along with the overlying retina from the remaining BM. The corrected result is shown in Fig. 9(b). We also refer to (Xiang et al., 2018; Scy et al., 2012; Staurenhi et al., 2014) for manual correction of mislabels in other diseases. More examples can be seen in Fig. 10.

Since OCTA-500 contains rich diseases, the layer labels are suitable to explore the retinal layer segmentation algorithm under the condition of disease diversity. As this paper tends to introduce the next CAVF task, we will not explore the layer segmentation algorithm further. For more research on the layer segmentation algorithm in OCTA-500, please refer to (Zhang et al., 2020; Yang et al., 2022), both of which use the data in OCTA-500.

## 4 CAVF Task and Baselines

### 4.1 CAVF Task

Based on the segmentation labels provided by our dataset, we propose a new CAVF task, which unifies Capillary segmentation, Artery segmentation, Vein segmentation, and FAZ segmentation under one framework. Fig. 10 provides some examples of the multi-object segmentation labels. Compared with segmenting them one by one, the proposed CAVF task will bring convenience in the computation of quantitative indicators and the evaluation of model performance:

- Understanding retinal diseases often requires quantitative analysis of different structural indicators (vessel density, FAZ area, etc.), and the calculation of these indicators often relies on different segmentation tasks. Previous studies (Ma et al., 2020; Díaz et al., 2019; Xu et al., 2022; Pissas et al., 2020; Lo et al., 2020) focused on single segmentation task. Training one model per task would be inconvenient for the application. Our proposed multi-object segmentation task unifies several interrelated segmentation tasks: the FAZ is surrounded by capillaries; arteries and veins are mutually exclusive, and one vessel cannot be identified as both an artery and a vein, etc. The proposed multi-object segmentation task allows one model to solve multiple segmentation problems, which reduces the computational burden and brings convenience to clinical applications.
- Multi-object segmentation aggregates the

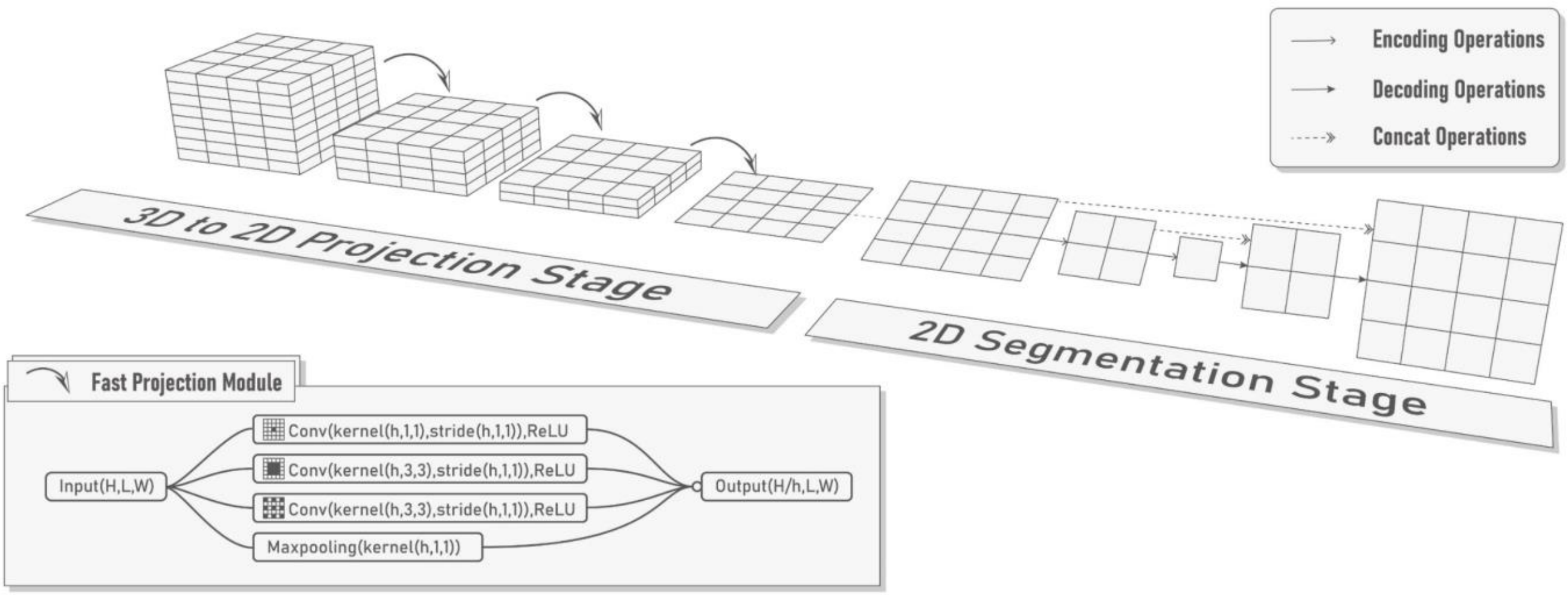

Fig. 11. Proposed IPN-V2 architecture for 3D-to-2D segmentation.

characteristics of each individual segmentation, allowing a more comprehensive evaluation of model performance. For example, capillary segmentation may require the model with the ability to maintain high-resolution information, FAZ segmentation may require the model to consider the location characteristics of objects, and artery-vein segmentation may require the model to extract higher-level features to distinguish them.

## 4.2 Baselines

To initially explore our proposed CAVF task, we selected several 2D-to-2D baselines as described in Section 4.2.1, and optimized the 3D-to-2D baseline as described in Section 4.2.2. Codes of these baselines are available in our dataset.

### *4.2.1 2D-to-2D Baselines*

The 2D-to-2D baselines we selected are U-Net (Ronneberger et al., 2015), UNet ++ (Zhou et al., 2018), UNet 3+ (Huang et al.,2020), Attention U-Net (Oktay et al., 2018), CS-Net (Mou et al., 2021), and AV-Net (Alam et al., 2020). Among them, U-Net is the most commonly used convolutional neural network in medical image segmentation, and it has been proven to be fast and accurate, even with few training images. Many recent OCTA segmentation methods (Mou et al., 2021; Guo et al., 2019a; Alam et al., 2020; Mou et al., 2019; Pissas et al., 2020; Lo et al., 2020; Li et al., 2020b) are extensions of U-Net and profit from the basic concepts of these methods. UNet ++, UNet 3+, and Attention U-Net optimize the structure and function based on U-Net, and they are also selected as baselines. Focusing on OCTA segmentation, we choose the baselines CS-Net and AV-Net, which are designed for vessel segmentation and artery-vein segmentation, respectively, and we tested their performance on our dataset. All of the above baselines take 2D projection images as input (the network inputs will be discussed in Section 5.2), and all hyperparameters are tuned to yield full segmentation performance.

### *4.2.2 3D-to-2D Baselines*

Our previous work (Li et al., 2020a) introduced an image projection network (IPN) for 3D to 2D segmentation, which inputs 3D OCT/OCTA volumes and outputs 2D segmentation results end-to-end. This 3D-to-2D segmentation brings several benefits: for example, the segmentation no longer needs to rely on layer segmentation to generate projection images, which avoids the failure of layer segmentation under disease conditions; it utilizes the complete 3D information, which reduces information loss and improves segmentation performance. To explore 3D-to-2D segmentation on the CAVF task, we use IPN as one of the baselines. However, it does not perform well in our experiments (see Section 5.2), as the CAVF task is more challenging than previous segmentation of large vessels or FAZ alone. Thus, in this work, we also introduce IPN-V2 (Fig. 11), which optimized the structure of the IPN, to serve as one of the baselines.

We first point out the limitations of the IPN in two aspects. First, the training and testing of the IPN takes up considerable GPU memory and is inefficient. Due to limited computing resources, we usually split the volume data (with a size of 640 × 400 × 400) into small blocks (e. g., with a size of 640 × 100 × 100) to train the IPN instead of inputting the whole volume data, which will lose some global information. Second, IPN has no down-sampling operation in the horizontal direction, so it lacks high-level semantic information, which is also one of the reasons for its poor performance on the proposed task.

To address the above limitation, we rethought the spatial structure of the IPN, as shown in Fig. 12. The 3D convolution operations with a kernel size of 3×3×3 in the first half of the network (Fig. 12, blue box) take up the most computational space. However, these convolutions in a large 3D space are inefficient: OCT/OCTA volumes contain a large amount of background, and it is not necessary to extract features in these positions. Thus, we consider first compressing the 3D information to 2D quickly to reduce the computational space of this part, and then further segment the target from the extracted 2D features. Based on the above considerations, the proposed IPN-V2 is thus designed and consists of two stages: the 3D-to-2D projection stage and the 2D segmentation stage.

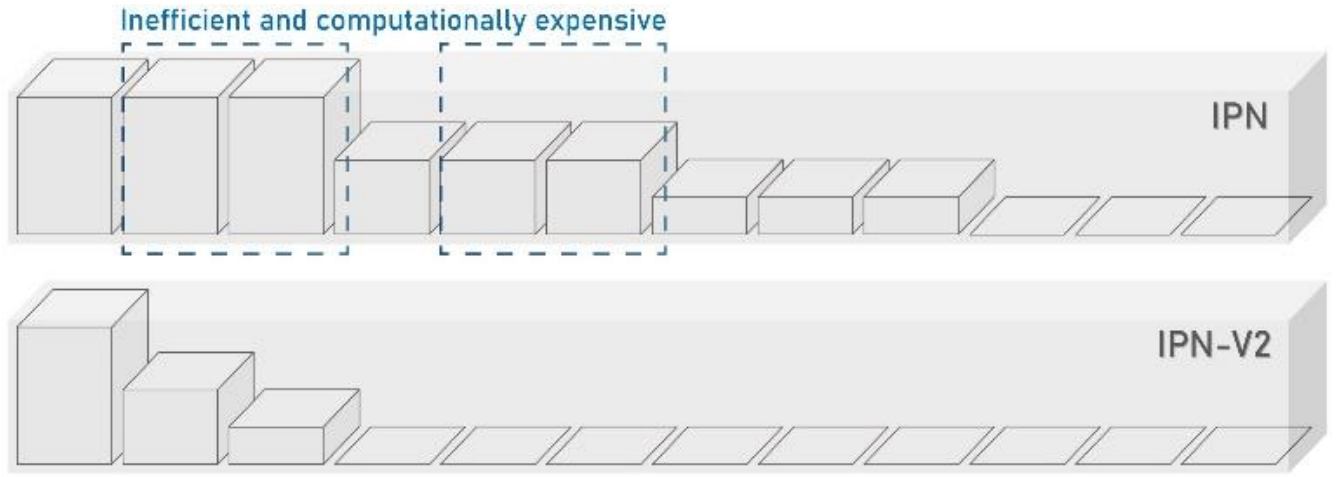

Fig. 12. Schematic representation of the computational space of IPN and IPN-V2.

**3D-to-2D projection stage:** This stage is mainly to quickly compress 3D information into a 2D space. To this end, we design the fast projection module (FPM). As shown in Fig. 11, the FPM includes four down-sampling operations, which are three different scales of convolution and the unidirectional pooling used in IPN. Multi-scale convolutions are used to extract and condense useful features, and unidirectional pooling can compress features and make training more stable. The FPM inputs the 3D volume with a size of (H, L, W) and the output volume with a size of (H/h, L, W), where h is the compression factor. Through several FPMs, 3D information is condensed into a 2D space to obtain a series of 2D features.

**2D segmentation stage:** The obtained 2D features are further segmented by a 2D segmentation network to achieve the complete segmentation task. In theory, the 2D segmentation network can use an arbitrary 2D convolutional neural network (CNN), such as the U-Net++ or other 2D baselines mentioned above. In this paper, to provide a valuable comparison, the 2D segmentation network is designed as a classic encoder-decoder structure similar to a U-Net. We also considered that IPN lacks high-level semantic information. The multi-scale semantic representation provided by this design just overcomes this limitation.

Note that, although IPN-V2 is designed to contain the above two stages, it is still an independent network that can be trained end-to-end. After the above optimizations, IPN-V2 can take the entire OCT/OCTA volume as input and achieve competitive segmentation on the CAVF task. In this paper, IPN-V2 is considered an important 3D-to-2D segmentation baseline. Ablation study of IPN-V2 see Supplementary Materials.

### 4.3 Evaluation Metrics

To objectively evaluate the segmentation performance of each object in our segmentation task, the following metrics can be calculated and compared:

1) Dice coefficient (Dice): 2TP/(2TP+FP+FN);
2) Intersection over union (IoU): TP/(TP+FP+FN);
3) Accuracy (ACC): (TP+TN)/(TP+FP+TN+FN);
4) Sensitivity (SE): TP/(TP+FN);
5) Specificity (SP): TN/(TN+FP);

where TP is true positive, FP is false positive, TN is true negative, and FN is false negative. To measure the comprehensive segmentation performance of the model, we use mean intersection over union (mIoU) as an important metric for the proposed CAVF task, and it is denoted as:

$$mIoU = \frac{1}{k}\sum_{i=1}^{k} IoU$$

where k is the number of segmentation objects. In addition, we also used the number of parameters (Params), the GPU memory occupied per batch (Memory) and test speed (reading and inference time for each subject) to reflect the computational complexity of the model.

## 5 Experiments

We quantitatively and qualitatively studied the CAVF task on the OCTA-500. We started with introducing the experimental setting in Section 5.1. Then, we evaluated the impact of several dataset characteristics: the training set size (Section 5.2), the model input (Section 5.3), the baselines (Section 5.4) and the diseases (Section 5.5).

### 5.1 Experimental Settings

We set the training set, validation set and test set on two subsets of OCTA-500. OCTA_6mm includes a training set (NO. 10001-NO. 10240), a validation set (NO. 10241-NO. 10250), and a test set (NO. 10251-NO. 10300). OCTA_3mm includes a training set (NO. 10301-NO. 10440), a validation set (NO. 10441-NO. 10450), and a test set (NO. 10451-NO. 10500). The training set is used to train network models; the validation set is used to select the best model; and the test set is used for evaluation.

All baselines are implemented with PyTorch on 2 NVIDIA GeForce RTX 3090 GPU. Cross entropy loss $L_{CE}$ and Dice loss $L_{DICE}$ are the two most popular loss functions in segmentation tasks, as described in (Ma et al., 2021). We use the unweighted sum $L_{CE} + L_{DICE}$ as the default loss function. We train the network using Adam optimization with a batch size of 4 and an initial learning rate of 0.0005. Each baseline iterates at least 300 epochs and saves the model every epoch. OCT/OCTA volumes are resized to 128 × 400 × 400 for OCTA_6mm and 128 × 304 × 304 for OCTA_3mm using bilinear interpolation. Other experimental settings (number of training sets, types of input images, etc.) will be given or discussed in the following sections.

### 5.2 Number of Training Data

We explore how the segmentation quality varies with increasing amount of training data. More specifically, we select the 2D-to-2D baseline U-Net and 3D-to-2D baseline IPN-V2 to perform experiments on the OCTA_6mm. The input of U-Net is all projection maps as introduced in Section 3.3, and the input of IPN-V2 is the OCTA volume. The training data are selected from the training set, and the number is set to 5, 10, 20, 40, 80, 120, 160, 200, and 240. The evaluation is performed on the test set of OCTA_6mm.

The results of this experiment are shown in Fig. 13. As it can be observed, the mIoU increases as we increase the number of training data samples. Specifically, when the number is less than 120, the mIoU increases rapidly. When the number is more than 120, the increase in mIoU becomes less pronounced, but we can still see some subtle improvements from the examples. In our opinion, the sample number of 120 is a basic requirement for this task, and the sample numbers of OCTA_6mm and OCTA_3mm are 300 and 200, respectively, which meet this requirement.

We also discuss the segmentation performance of each object in this experiment. From Fig. 13, we can see that the IoUs of artery, vein and FAZ show roughly similar trends to the mIoU, while the obvious difference is that the number of samples required for capillary segmentation using U-Net is very small, indicating that the capillary segmentation is easier than the segmentation of arteries, veins and FAZ. Additionally, FAZ segmentation appears to require the highest number of training samples, which may be related to the morphological diversity of the FAZ.

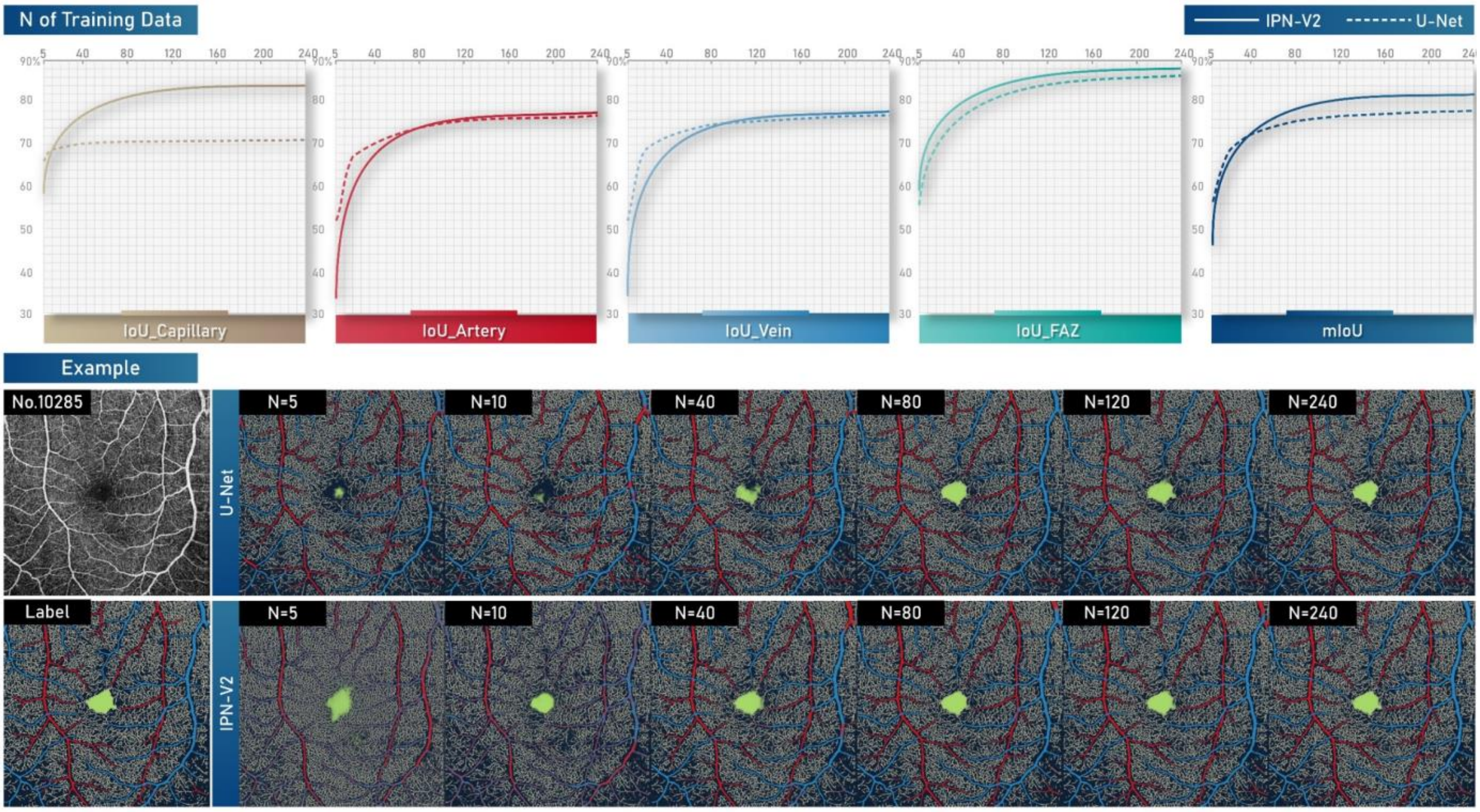

Fig. 13. Evaluations and examples in the experiment of varying training data size.

**TABLE 2**
The effect of different inputs on segmentation performance (IoU) using OCTA_6mm.

| Methods | Input | IoU (%) | | | | mIoU |
|---|---|---|---|---|---|---|
| | | C | A | V | F | (%) |
| U-Net | B5 | 69.05 | 75.49 | 76.68 | 84.99 | 76.55 |
| U-Net | B1-B3 | 27.01 | 63.62 | 65.21 | 69.92 | 56.44 |
| U-Net | B4-B6 | 71.00 | 75.70 | 76.51 | **86.69** | 77.47 |
| U-Net | B1-B6 | **71.01** | **75.94** | **77.29** | 86.04 | **77.57** |
| IPN-V2 | OCT | 34.93 | 68.38 | 70.13 | 68.90 | 60.59 |
| IPN-V2 | OCTA | **84.34** | 76.74 | 77.26 | **88.76** | **81.77** |
| IPN-V2 | Both | 80.66 | **76.78** | **77.79** | 87.75 | 80.75 |

U-Net can be observed to perform better than IPN-V2 when the number of samples is small. When the number is more than 40, IPN-V2 is better. The above observations indicate that the 3D-to-2D method requires more data than the 2D-to-2D method. Interestingly, we can also observe that when the number is more than 20, the IoU of the capillary segmentation of IPN-V2 is significantly higher than that of U-Net because the noise in the 2D projection map is easily misclassified as capillaries, while in 3D volumes this noise does not constitute a complete vascular structure and is easily distinguished.

### 5.3 Types of Input Data

Since different types of input data have differentiated information, we conducted a set of experiments to understand the effectiveness of using different inputs. In this section, we still selected the U-Net and IPN-V2 to perform experiments on the OCTA_6mm. The input settings considered include OCT projection maps B1-B3, OCTA projection maps B4-B6, OCT volumes, and OCTA volumes.

Table 2 shows the results of these experiments. It can be observed that the best input for 2D-to-2D segmentation is all B1-B6 projection maps, and the mIoU with B4-B6 projection maps as input is only 0.1 lower than it, indicating that the OCTA projection maps provide the dominant information. For 3D-to-2D segmentation, the highest mIoU is with only the OCTA volume input. It is difficult to segment capillaries using only OCT volume as input (IoU = 34.93 %), and adding OCT volume has a certain inhibition on capillary segmentation. Based on the above results, we use projection maps B1-B6 as default input for the 2D-to-2D baselines and OCTA volumes as default input for the 3D-to-2D baselines.

### 5.4 Comparison of Different Baselines

In this section, we compare the performance of all baseline methods on the CAVF segmentation task using two subsets of OCTA-500. All the baseline networks are trained using the entire training set. We used the validation set for model selection. Table 3 shows the quantitative results (IoU) of all baselines evaluated on the test set. We can conclude that U-Net, UNet++, UNet 3+ and Attention U-Net perform better than AV-Net and CS-Net, possibly because AV-Net and CS-Net are designed for a single task and thus do not generalize well on our proposed task. In the 2D baselines,

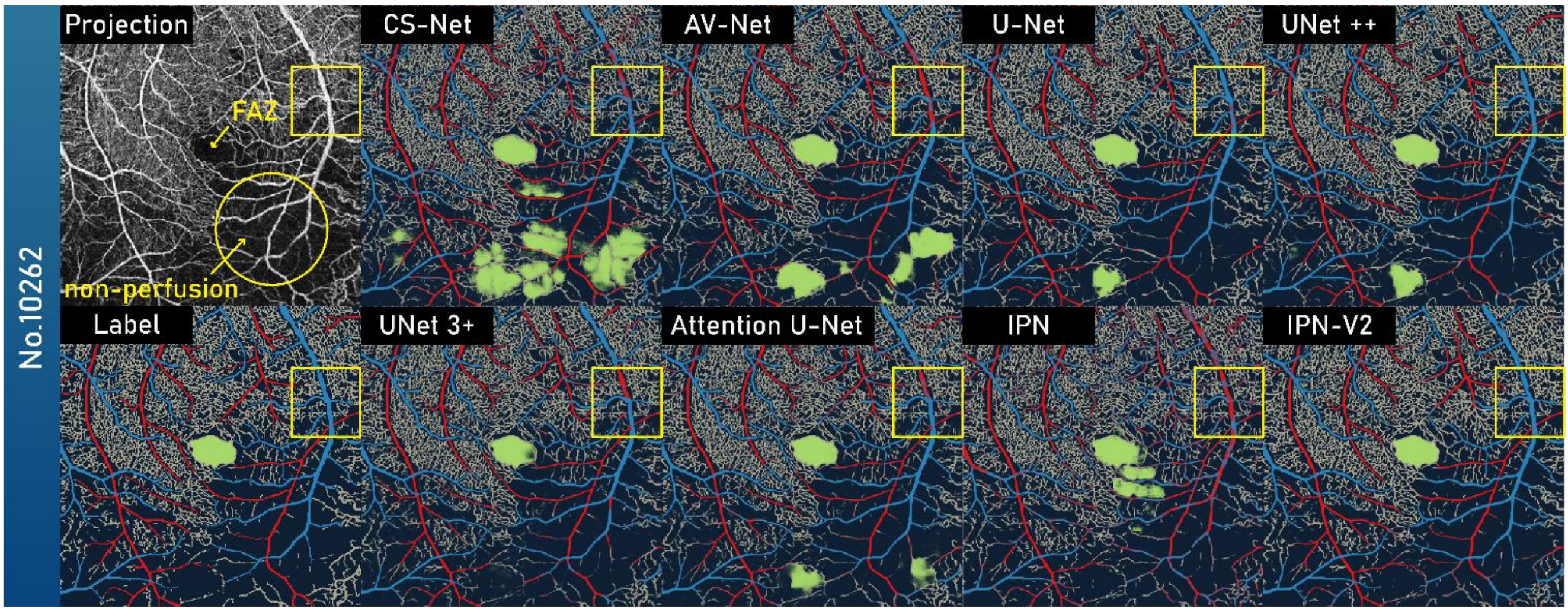

Fig. 14. An example of segmentation results using different baselines.

**TABLE 3**
Results (IoU) of different baselines on OCTA-500.

| Methods | OCTA_6mm test set (No. 10251-No.100300) | | | | | OCTA_3mm test set (No. 10451-No.10500) | | | | |
|---|---|---|---|---|---|---|---|---|---|---|
| | IoU (%) | | | | mIoU (%) | IoU (%) | | | | mIoU (%) |
| | C | A | V | F | | C | A | V | F | |
| CS-Net | 69.10 | 64.72 | 67.61 | 78.30 | 69.93 | 77.02 | 72.15 | 68.76 | 92.55 | 77.62 |
| AV-Net | 70.49 | 73.69 | 74.86 | 81.63 | 75.17 | 77.87 | 69.24 | 63.29 | 94.31 | 76.18 |
| U-Net | 71.01 | 75.94 | 77.29 | **86.04** | 77.57 | 78.63 | 78.95 | 78.32 | 95.24 | 82.78 |
| UNet ++ | 71.12 | 76.03 | **77.66** | 85.79 | **77.65** | 78.63 | 80.91 | 79.42 | 95.40 | 83.59 |
| UNet 3+ | 71.07 | 75.31 | 76.94 | 83.56 | 76.72 | 78.55 | 80.88 | 79.43 | 94.75 | 83.40 |
| Attention U-Net | **71.51** | **76.14** | 77.54 | 84.95 | 77.53 | **78.78** | **81.46** | **80.07** | **95.81** | **84.03** |
| IPN | 79.82 | 59.92 | 60.18 | 79.31 | 69.81 | 83.64 | 71.15 | 67.81 | 91.61 | 78.55 |
| IPN-V2 | **84.34** | **76.74** | **77.26** | **88.76** | **81.77** | **86.16** | **82.26** | **81.38** | **95.15** | **86.24** |

Attention U-Net performs well, and achieves the best IoU for all categories in the OCTA_3mm, indicating that the attention strategy may improve the segmentation results of this task. In all baselines, IPN-V2 achieves the best mIoU on both OCTA_6mm and OCTA_3mm. IPN-V2 outperforms other models by a large margin on capillary segmentation and achieves competitive results on artery segmentation, vein segmentation, and FAZ segmentation.

Fig. 14 shows an example of segmentation results using different baselines. This example comes from a DR patient. We can observe that the methods other than IPN-V2 show varying degrees of over-segmentation and under-segmentation in FAZ segmentation. DR disease is often accompanied by non-perfusion of capillary. The grayscale range of the non-perfusion in the projection image is similar to that of the FAZ, which is prone to mis-segmentation. The segmentation result using IPN-V2 is close to the manual label, indicating that the use of 3D information may reduce this interference. The results of artery and vein segmentation have inconsistencies on different segments of the same vessel (marked by yellow box). IPN-V2 performs better than other methods in this example. However, IPN-V2 does not completely avoid this issue. This issue is still a challenge in artery-vein segmentation.

More evaluations, including speed, are provided in the Supplementary Materials.

### 5.5 Performance in Different Diseases

To evaluate the segmentation performance under different disease conditions, we performed a 3-fold cross-validation on OCTA_6mm using the IPN-V2 model to obtain the segmentation results of all subjects. We further counted the quantitative metrics of these results according to the category of the diseases. Fig. 15 shows the results and examples of this experiment. The ranking by mIoU is normal, CSC, CNV, AMD, DR and RVO. The segmentation results of the normal subjects achieve the best mIoU and the best IoUs of all subtasks showing that CAVF segmentation in normal retinas can achieve reliable results. The segmentation performance in the disease case is relatively poor. In particular, AMD, DR, and RVO are often accompanied by non-perfusion and vascular morphological changes, which increase the difficulty of segmentation. From the examples of these diseases, we can also observe the over-segmentation and under-segmentation of the FAZ, as well as the mis-segmentation of some arteries and veins. Therefore, there is still a need to further improve the segmentation performance under disease conditions in the future.

## 6 Discussion

We have demonstrated the CAVF task to achieve joint segmentation of capillaries, arteries, veins, and the FAZ.

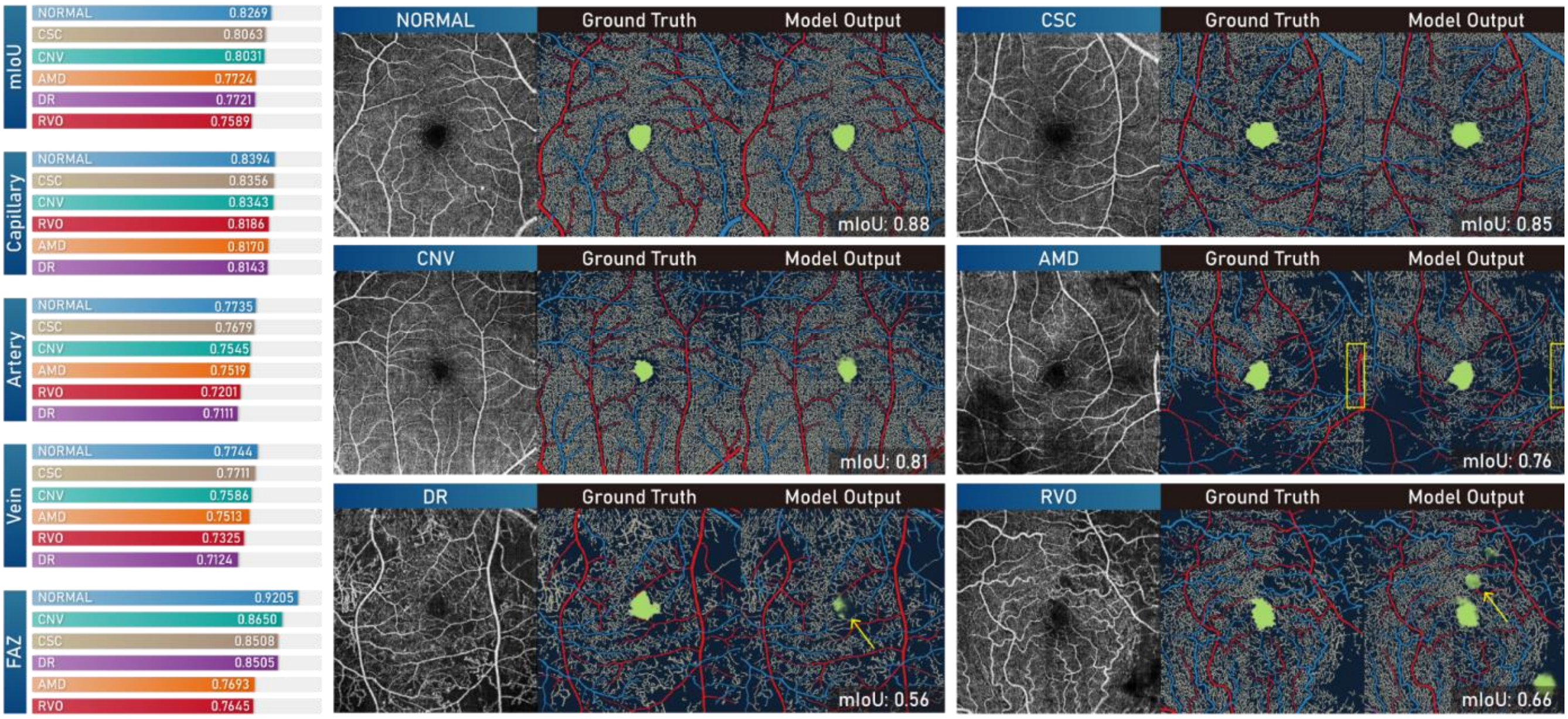

Fig. 15. Segmentation performance on different diseases using IPN-V2 on OCTA_6mm.

Segmentation labels in OCTA-500 also allow us to achieve large vessel segmentation, 3D FAZ segmentation, and layer segmentation. Experiments for each single segmentation task are provided in the Supplementary Materials. These tasks are not only used to optimize and evaluate algorithms, but more importantly, they provide diverse retinal image quantification and evaluation methods, which will play an important role in disease analysis. For example, layer segmentation is used to calculate retinal layer thickness (Goebel et al., 2002), vessel segmentation is used to calculate vessel density (Livia et al., 2019), FAZ segmentation is used to calculate FAZ area (Samara et al., 2015), etc.

OCTA-500 also provides a variety of disease labels, which can be used for disease classification. A recent study (Lin et al., 2021) discussed the classification performance of normal, DR and AMD using the projection images in OCTA-500. Still, more diseases need to be discussed, and 3D OCT/OCTA volumes have not been used. OCTA-500 is one of the few datasets that contains paired OCT/OCTA volumes. The multi-modality data will support a wider range of research, such as modality transformation (Lee et al., 2019), image denoising (Yang et al., 2020), artifact removal (Zhang et al., 2017), multi-modality fusion (Jia et al.,2014), etc. We also expect OCTA-500 to stimulate more interesting research topics.

## 7 Conclusion

We have introduced the new OCTA-500 dataset, which contains OCTA imaging from 500 subjects and provides a rich set of images and annotations. Based on the provided segmentation annotations, we have proposed a new CAVF segmentation task that integrates artery segmentation, vein segmentation, capillary segmentation, and FAZ segmentation under a unified framework. Focusing on the proposed CAVF task, we optimized the 3D-to-2D network IPN to IPN-V2 to serve as one of the baselines. We have explored the effect of several dataset characteristics on the CAVF task: the training set size, the model input, the baselines, and the diseases.

The experiments show that data are a driving factor for the segmentation performance in the proposed task. Our dataset has been at a reasonable level in terms of data scale. The proposed IPN-V2 has improved the quality and speed of segmentation by a large margin compared with IPN, and achieved competitive results. We also show that the disease diversity of OCTA-500 increases the challenge of the segmentation task. The considered deep learning methods have not yet been saturated in this task. Future improvement will come from better methods and increased data.

OCTA-500 allows us to implement a variety of segmentation tasks, which will provide a systematic quantitative framework for retinal image analysis. We also discussed its potential applications in other OCTA studies. Hence, we expect that it will stimulate research toward the quantification, analysis and application of OCT/OCTA images. Our future plans are to continue collecting images, annotating ground truths, and optimizing methods for more OCTA studies.

## Acknowledgment

This study was supported by National Natural Science Foundation of China (62172223, 61671242), and the Fundamental Research Funds for the Central Universities (30921013105).

## References

Agarwal, A., Balaji, J. J., Raman, R., Lakshminarayanan, V., 2020. The foveal avascular zone image database (FAZID). Applications of Digital Image Processing XLIII, Proc. SPIE 11510.

Aharony, O., Gal-Or, O., Polat, A., Nahum, Y., Weinberger, D., Zimmer, Y., 2019.

Automatic characterization of retinal blood flow using OCT angiograms. Translational vision science & technology 8(4), 1-10.

Alam, M., Thapa, D., Lim, J. I., Cao, D., Yao, X., 2017. Quantitative characteristics of sickle cell retinopathy in optical coherence tomography angiography. Biomedical Optics Express 8(3), 1741-1753.

Alam, M., Toslak, D., Lim, J. I., Yao, X., 2018. Color fundus image guided artery-vein differentiation in optical coherence tomography angiography. Translational Vision Science & Technology 59(12), 4953-4962.

Alam, M., Lim, J. I., Toslak, D., Yao, X., 2019a. Differential artery-vein analysis improves the performance of octa staging of sickle cell retinopathy. Translational Vision Science & Technology, 8(2),1-8.

Alam, M., Toslak, D., Lim, J. I., Yao, X., 2019b. Oct feature analysis guided artery-vein differentiation in OCTA. Biomedical Optics Express 10(4), 2055-2066.

Alam, M., Le, D., Son, T., Lim, J. I., Yao, X., 2020. AV-Net: deep learning for fully automated artery-vein classification in optical coherence tomography angiography. Biomedical Optics Express 11(9), 5249-5257.

Antony, B., Abramoff, M. D., Tang, L., Ramdas, W. D., Vingerling, J. R., Jansonius, N. M., Sonka, M., Garvin, M. K., et al., 2011 Automated 3-d method for the correction of axial artifacts in spectral-domain optical coherence tomography images. Biomedical Optics Express 2(8), 2403-2416.

Balaratnasingam, C., Inoue, M., Ahn, S., Mccann, J., Dhrami-Gavazi, E., Yannuzzi, L. A., Freund, K. B., 2016. Visual acuity is correlated with the area of the foveal avascular zone in diabetic retinopathy and retinal vein occlusion. Ophthalmology 123(11), 2352-2367.

Bogunovic, H., Venhuizen, F., Klimscha, S., Apostolopoulos, S., Bab-Hadiashar, A. and Schmidt-Erfurth, U., et al., 2019. Retouch -the retinal oct fluid detection and segmentation benchmark and challenge. IEEE Trans. Med. Imaging 38(8), 1858-1874.

Chen, Q. and Niu, S., 2016. High-low reflectivity enhancement based retinal vessel projection for SD-OCT images. Medical Physics 43(10), 5464-5474.

Chu, Z., Lin, J., Gao, C., Xin, C., Zhang, Q., Chen, C., Roisman, L., Gregori, G., Rosenfeld, P. J., Wang, R. K., 2016. Quantitative assessment of the retinal microvasculature using optical coherence tomography angiography. Journal of Biomedical Optics 21(6), 066008.

Cicek, O., Abdulkadir, A., Lienkamp, S. S., Brox, T., Ronneberger, O., 2016. 3D U-Net: Learning dense volumetric segmentation from sparse annotation. International Conference on Medical Image Computing and Computer-Assisted Intervention 2016.

Conrath, J., Giorgi, R., Raccah, D., Ridings, B., 2005. Foveal avascular zone in diabetic retinopathy: quantitative vs qualitative assessment. Eye 19(3), 322-326.

Deng, J., Dong, W., Socher, R., Li, L., Li, K., Li, F., 2009. ImageNet: A large-scale hierarchical image database. IEEE Conf. Vis. Pattern Recognit., 248-255.

Díaz, M., Novo, J., Cutrín, P., Ulla, F.G., Penedo, M. G., Ortega, M., 2019. Automatic segmentation of the foveal avascular zone in ophthalmological OCT-A images. Plos One 14(2), 1–22.

Eladawi, N., Elmogy, M., Helmy, O., Aboelfetouh, A., Riad, A., Sandhu, H., Schaal, S., El-Baz, A., 2017. Automatic blood vessels segmentation based on different retinal maps from octa scans. Computers in Biology and Medicine 89, 150-161.

Frangi, A. F., Niessen, W. J., Vincken, K. L., Viergever, M. A., 1998. Multiscale vessel enhancement filtering. International Conference on Medical Image Computing and Computer-Assisted Intervention 1998, 130-137.

Fraz, M. M., Remagnino, P., Hoppe, A., Uyyanonvara, B., Rudnicka, A. R., Owen, C. G., Barman, S. A., 2012. Blood vessel segmentation methodologies in retinal images – a survey. Computer Methods & Programs in Biomedicine 108(1), 407-433.

Freiberg, F. J., Pfau, M., Wons, J., Wirth, M. A., Becker, M. D., Michels, S., 2016. Optical coherence tomography angiography of the foveal avascular zone in diabetic retinopathy. Graefes Arch. Clin. Exp. Ophthalmol. 254(6), 1051-1058.

Garvin, M. K., Abramoff, M. D., Wu, X., Russell, S. R., Burns, T. L., Sonka, M., 2009. Automated 3-d intraretinal layer segmentation of macular spectral-domain optical coherence tomography images. IEEE Trans. Med. Imaging 28(9), 1436-1447.

Giarratano, Y., Bianchi, E., Gray, C., Morris, A., MacGillivray, T., Dhillon B., Bernabeu, M. O., 2020. Automated segmentation of optical coherence tomography angiography images: benchmark data and clinically relevant metrics. Translational vision science & technology 9(13), 1-10.

Goebel, W., Kretzchmar-Gross, T., 2002. Retinal thickness in diabetic retinopathy: a study using optical coherence tomography (OCT). Retina 22(6), 759-767.

Guo, M., Zhao, M., Cheong, A. M. Y., Dai, H., Lam, A. K. C., Zhou, Y., 2019b. Automatic quantification of superficial foveal avascular zone in optical coherence tomography angiography implemented with deep learning. Visual Computing for Industry Biomedicine and Art 2, 1-9.

Guo, M., Zhao, M., Cheong, A. M., Corvi, F., Chen, X., Chen, S., Zhou, Y., Lam, A. K., 2021. Can deep learning improve the automatic segmentation of deep foveal avascular zone in optical coherence tomography angiography. Biomedical Signal Processing and Control 66, 102456.

Guo, Y., Camino, A., Wang, J., Huang, D., Hwang, T. S., Jia, Y., 2018. MEDnet, a neural network for automated detection of avascular area in oct angiography. Biomedical Optics Express 9(11), 5147-5158.

Guo, Y., Hormel, T. T., Xiong, H., Wang, B., Camino, A., Wang, J., Huang, D., Hwang, T. S., Jia, Y., 2019a. Development and validation of a deep learning algorithm for distinguishing the nonperfusion area from signal reduction artifacts on OCT angiography. Biomedical Optics Express 10(7), 3257-3268.

Hoover, A. D., Kouznetsova, V., Goldbaum, M., 2002. Locating blood vessels in retinal images by piecewise threshold probing of a matched filter response. IEEE Trans. Med. Imaging 19(3), 203-210.

Hormel, T. T., Wang, J., Bailey, S. T., Hwang, T. S., Huang, D., Jia, Y., 2018. Maximum value projection produces better en face oct angiograms than mean value projection. Biomedical Optics Express 9(12), 6412-6424.

Huang, D., Swanson, E. A., Lin, C. P., Schuman, J. S., Stinson, W. G., Chang, W., Hee, M. R., Flotte, T., Gregory, K., Puliafrro, C. A., Fujimoto, J. G., 1991. Optical coherence tomography. Science 254 (5035), 1178-1181.

Huang, H., Lin, L., Tong, R., Hu, H., Zhang, Q., Iwamoto, Y., et al., 2020 UNet 3+: A full-scale connected UNet for medical image segmentation. IEEE International Conference on Acoustics, Speech and Signal Processing 2020.

Jabour, C., Garcia, D., Mathis, T., Loria, O., Rochepeau, C., Harbaoui, B., Lantelme, P., Vray, D., Merveille, O., 2021. Robust foveal avascular zone segmentation and anatomical feature extraction from OCT-A handling inter-expert variability. IEEE 18th International Symposium on Biomedical Imaging 2021.

Jia, Y., Tan, O., Tokayer, J., Potsaid, B., Wang, Y., Liu, J. J., Kraus, M. F., Subhash, H., Fujimoto, J. G., Hornegger, J., Huang, D., 2012. Split-spectrum amplitude-decorrelation angiography with optical coherence tomography. Optics Express 20(4), 4710-4725.

Jia, Y., Bailey, S. T., Wilson, D. J., Tan, O., Klein, M. L., Flaxel, C. J., Potsaid, B., Liu, J. J., Lu, C. D., Kraus, M. F., Fujimoto, J. G., Huang, D., 2014. Quantitative optical coherence tomography angiography of choroidal neovascularization in age-related macular degeneration. Ophthalmology 121(7), 1435-1444.

Jiang, H., Wei, Y., Shi, Y., Wright, C. B., Sun, X., Gregori, G., Zheng, F., Vanner, E. A., Lam, B. L., Rundek, T., Wang, J., 2018. Altered macular microvasculature in Mild Cognitive impairment and Alzheimer disease. J Neuro-Ophthalmology Society 38(3), 292-298.

Kashani, A. H., Chen, C. L., Gahm, J. K., Zheng, F., Richter, G. M., Rosenfeld, P. J., Shi, Y., Wang, R. K., 2017. Optical coherence tomography angiography:

a comprehensive review of current methods and clinical applications. Progress in Retinal and Eye Research 60, 66-100.

Kondermann, C., Kondermann, D., Yan, M., 2007. Blood vessel classification into arteries and veins in retinal images. SPIE, doi: 10.1117/12.708469.

Kwapong, W. R., Ye, H., Peng, C., Zhuang, X., Wang, J., Shen, M., Lu, F., 2018. Retinal microvascular impairment in the early stages of Parkinson's disease. Investigative Ophthalmology & Visual Science 59, 4115-4122.

Laíns, I., Wang, J. C., Cui, Y., Katz, R., Vingopoulos, F., Staurenghi, G., Vavvas, D. G., Miller, J. W., Miller, J. B., 2021. Retinal applications of swept source optical coherence tomography (OCT) and optical coherence tomography angiography (OCTA). Progress in Retinal and Eye Research 64, 100951.

Lavia, C., Bonnin, S., Maule, M., Erginay, A., Tadayoni, R. and Gaudric, A., 2019. Vessel density of superficial, intermediate, and deep capillary plexuses using optical coherence tomography angiography. Retina 39, 247-258.

Lee, C. S., Tyring, A. J., Wu, Y., Xiao, S., Rokem, A. S., et al., 2019. Generating retinal flow maps from structural optical coherence tomography with artificial intelligence. Scientific Reports 9(1), 5694.

Li, K., Wu, X., Chen, D. Z., Sonka, M., 2006. Optimal surface segmentation in volumetric images - a graph-theoretic approach. IEEE Trans. Pattern Anal. Mach. Intell. 28(1), 119-134.

Li, A., You, J., Du, C., Pan, Y., 2017. Automated segmentation and quantification of oct angiography for tracking angiogenesis progression. Biomedical Optics Express 8(12), 5604-5616.

Li, M., Chen, Y., Ji, Z., Xie, K., Yuan, S., Chen, Q., Li, S., 2020a. Image projection network: 3D to 2D image segmentation in OCTA images. IEEE Trans. Med. Imaging 39(11), 3343-3354.

Li, M., Wang, Y., Ji, Z., Fan, W., Yuan, S., Chen, Q., 2020b. Fast and robust fovea detection framework for OCT images based on foveal avascular zone segmentation. OSA Continuum 3(3), 528-541.

Li, M., Zhang, W., Chen, Q., 2022a. Image magnification network for vessel segmentation in OCTA images. Chinese Conference on Pattern Recognition and Computer Vision 2022. *arXiv:2110.13428*.

Li, M., Huang, K., Zhang, Z., Ma, X., Chen, Q., 2022b. Label adversarial learning for skeleton-level to pixel-level adjustable vessel segmentation. arXiv: 2205.03646.

Li, W., Zhang, H., Li, F., Wang, L., 2022c. RPS-Net: An effective retinal image projection segmentation network for retinal vessels and foveal avascular zone based on OCTA data. Medical Physics 49(6), 3830-3844.

Liang, Z., Zhang, J., An, C., 2021. Foveal avascular zone segmentation of OCTA images using deep learning approach with unsupervised vessel segmentation. International Conference on Acoustics, Speech and Signal Processing 2021.

Lin, A., Fang, D., Li, C., Cheung, C. Y., Chen, H., 2020. Improved automated foveal avascular zone measurement in cirrus optical coherence tomography angiography using the level sets macro. Translational Vision Science & Technology 9(12), 1-10.

Lin, L., Wang, Z., Wu, J., Huang, Y., Lyu, J., Cheng, P., Wu, J., Tang, X., 2021. BSDA-Net: a boundary shape and distance aware joint learning framework for segmenting and classifying OCTA images. International Conference on Medical Image Computing and Computer-Assisted Intervention 2021.

Liu, Y., Carass, A., Zuo, L., et al., 2022. Disentangled representation learning for OCTA vessel segmentation with limited training data. IEEE Trans. Med. Imaging 41(12), 3686-3698.

Lo, J., Heisler, M., Vanzan, V., Karst, S., Matovinovic, I. Z., Loncaric, S., Navajas, E. V., Beg, M. F., Sarunic, M. V., 2020. Microvasculature segmentation and intercapillary area quantification of the deep vascular complex using transfer learning. Translational vision science & technology 9(2), 1-12.

Lu, Y., Simonett, J. M., Wang, J., Zhang, M., Hwang, T., Hagag, A. M., Huang, D., Li, D., Jia, Y., 2018. Evaluation of automatically quantified foveal avascular zone metrics for diagnosis of diabetic retinopathy using optical coherence tomography angiography. Investigative Ophthalmology & Visual Science 59(6), 2212-2221.

Ma, J., 2021. Cutting-edge 3d medical image segmentation methods in 2020: are happy families all alike? arXiv:2101.00232.

Ma, Y., Hao, H., Xie, J., Fu, H., Zhang, J., Yang, J., Wang, Z., Liu, J., Zheng, Y., Zhao, Y., 2020. ROSE: A retinal OCT-Angiography vessel segmentation da-taset and new model. IEEE Trans. Med. Imaging 40(3), 928-939.

Maio, L. G. D., Montorio, D., Peluso, S., Dolce, P., Salvatore, E., Michele, G. D., Cennamo, G., 2021. Optical coherence tomography angiography findings in Huntington's disease. Neurological Sciences42, 995-1001.

Mehta, N., Braun, P. X., Gendelman, I., Alibhai, A. Y., Arya, M., Duker, J. S., Waheed, N. K., 2020. Repeatability of binarization thresholding methods for optical coherence tomography angiography image quantification. Scientific Reports 10 (1), 15368.

Meiburger, K. M., Salvi, M., Rotunno, G., Drexler, W., Liu, M., 2021. Automatic segmentation and classification methods using optical coherence tomography angiography (OCTA): a review and handbook. Applied sciences 11(9734), 1-28.

Mirshahi, R., Anvari, P., Riazi-Esfahani, H., Sardarinia, M., Naseripour, M., Falavarjani, K. G., 2021. Foveal avascular zone segmentation in optical coherence tomography angiography images using a deep learning approach. Scientific Reports 11, 1031.

Moccia, S., Momi, E. D., Hadji, S. E., Mattos, L. S., 2018. Blood vessel segmentation algorithms — Review of methods, datasets and evaluation metrics. Computer Methods & Programs in Biomedicine158, 71-91.

Mookiah, M. R. K., Hogg, S., MacGillivray, T. J., Prathiba, V., Pradeepa, R., Trucco, E., et al., 2021. A review of machine learning methods for retinal blood vessel segmentation and artery/vein classification. Medical Image analysis 68, 101905.

Mou, L., Zhao, Y., Chen, L., Cheng, J., Gu, Z., Hao, H., Qi, H., Zheng, Y., Frangi, A., Liu, J., 2019. CS-Net: Channel and spatial attention network for curvilinear structure segmentation. International Conference on Medical Image Computing and Computer-Assisted Intervention 2019, 721-730.

Mou, L., Zhao, Y., Fu, H., Liu, Y., Cheng, J., Zheng, Y., Su, P., Yang, J., Chen, L., Frangi, A., Akiba, M., Liu, J., 2021. $CS^2$-Net: Deep learning segmentation of curvilinear structures in medical imaging. Medical Image Analysis 67, 101874.

Novosel, J., Vermeer, K. A., Jong, J. H. D., Wang, Z., Vliet, L. J. V., 2017. Joint segmentation of retinal layers and focal lesions in 3-d oct data of topologically disrupted retinas. IEEE Trans. Med. Imaging 36(6), 1276-1286.

Oktay, O., Schlemper, J., Folgoc, L. L., Lee, M., Heinrich, M., Misawa, K., Mori, K., Rueckert, D., et al., 2018. Attention u-net: learning where to look for the pancreas. International Conference on Medical Imaging with Deep Learning 2018.

Otsu, N., 1979. A threshold selection method from gray-level histograms. IEEE Transactions on Systems Man & Cybernetics 9(1), 62-66.

Owen, C. G., Rudnicka, A. R., Mullen, R., Barman, S. A., Monekosso, D., Whincup, P. H., Ng, J., Paterson, C., 2009. Measuring retinal vessel tortuosity in 10-year-old children: validation of the computer-assisted image analysis of the retina (caiar) program. Investigative Ophthalmology & Visual Science 50(5), 2004-2010.

Peng, L., Lin, L., Cheng, P., Wang, Z., Tang, X., 2021. FARGO: A joint framework for FAZ and RV segmentation from OCTA images. OMIA 2021, 1-10.

Pissas, T., Bloch, E., Cardoso, M. J., Flores, B., Georgiadis, O., Jalali, S., Ravasio, C., Stoyanov, D., Cruz, L. D., Bergeles, C., 2020. Deep iterative vessel segmentation in OCT angiography. Biomedical Optics Express 11(5), 2490-2509.

Prentasic, P., Heisler, M., Mammo, Z., Lee, S., Merkur, A., Navajas, E., Beg, M. F., Sarunic, M., Loncaric, S., 2016. Segmentation of the foveal microvasculature using deep learning networks. Journal of biomedical optics 21(7), 1-7.

Robbins, C. B., Grewal, D. S., Thompson, A. C., Soundararajan, S., Yoon, S. P., Polascik, B. W., Scott, B. L., Fekrat, S., 2022. Identifying peripapillary radial capillary plexus alterations in Parkinson's disease using OCT angiography. Ophthalmology Retina 6(1), 29-36.

Ronneberger, O., Fischer, P., Brox, T., 2015. U-Net: Convolutional networks for biomedical image segmentation. International Conference on Medical Image Computing and Computer-Assisted Intervention 2015.

Sakata, L. M., Deleon-Ortega, J., Sakata, V., Girkin, C. A., 2009. Optical coherence tomography of the retina and optic nerve – a review. Clinical & Experimental Ophthalmology 37, 90-99.

Samara, W. A., Say, E. A. T., Khoo, C. T. L., Higgins, T. P., Magrath, G., Ferenczy, S., et al., 2015. Correlation of foveal avascular zone size with foveal morphology in normal eyes using optical coherence tomography angiography. Retina 35(11), 2188-2195.

Spaide, R. F., Fujimoto, J. G., Waheed, N. K., Sadda, S. R., Staurenghi, G., 2018. Optical coherence tomography angiography. Progress in Retinal and Eye Research 64, 1-55.

Staal, J., Abramoff, M. D., Viergever, M. A., Ginneken, B. V., 2004. Ridge-based vessel segmentation in color images of the retina. IEEE Trans. Med. Imaging 23(4), 501-509.

Staurenghi, G., Sadda, S., Chakravarthy, U., Spaide, R. F., 2014. Proposed lexicon for anatomic landmarks in normal posterior segment spectral-domain optical coherence tomography: the in oct consensus Ophthalmology 121(8), 1572-1578.

Stefan, S., Lee, J., 2020. Deep learning toolbox for automated enhancement, segmentation, and graphing of cortical optical coherence tomography microangiograms. Biomedical Optics Express 11(12), 7325-7342.

Syc, S. B., Saidha, S., Newsome, S. D., Ratchford, J. N., Levy, M., Calabresi, P. A., et al., 2012. Optical coherence tomography segmentation reveals ganglion cell layer pathology after optic neuritis. Brain A Journal of Neurology 135, 521-533.

Terheyden, J. H., Wintergerst, M. W. M., Falahat, P., Berger, M., Holz, F. G., Finger, R. P., 2019. Automated thresholding algorithms outperform manual thresholding in macular optical coherence tomography angiography image analysis. Plos One 15(3), e0230260.

Vogl, W., Waldstein, S. M., Gerendas, B. S., Schmidt-Erfurthm U., Langs, G., 2017. Predicting macular edema recurrence from spatio-temporal signatures in optical coherence tomography images. IEEE Trans. Med. Imaging 36(9), 1773-1783.

Wang, X., Han, Y., Sun, G., Yang, F., Liu, W., Luo, J., Cao, X., Yin, P., Myers, F. L., Zhou, L., 2021. Detection of the microvascular changes of diabetic retinopathy progression using optical coherence tomography angiography. Translational vision science & technology, 10(7), 1-9.

Wu, Z., Wang, Z., Zou, W., Ji, F., Dang, H., Zhou, W., Sun, M., 2021. PAENet: a progressive attention-enhanced network for 3d to 2d retinal vessel segmentation. IEEE International Conference on Bioinformatics and Biomedicine 2021.

Xiang, D., Tian, H., Yang, X., Shi, F., Zhu, W., Chen, H., Chen, X., 2018. Automatic segmentation of retinal layer in OCT images with choroidal neovascularization. IEEE Trans. Image Processing 27(12), 5880-5891.

Xu, Q., Zhang, W., Zhu, H., Chen, Q., 2021. Foveal avascular zone volume: a new index based on optical coherence tomography angiography images. Retina 41(3), 595-601.

Xu, Q., Li, M., Pan, N., Chen, Q., Zhang, W., 2022 Priors-guided convolutional neural network for 3D foveal avascular zone segmentation. Optics Express 30(9), 14723-14736.

Xu, X., Chen, C., Ding, W., Yang, P., Lu, H., Xu, F., Lei, J., 2019. Automated quantification of superficial retinal capillaries and large vessels for diabetic retinopathy on optical coherence tomographic angiography. Journal of Biophotonics 12(11), e201900103.

Yang, J., Hu, Y., Fang, L., Cheng, J., Liu, J., 2020. Universal digital filtering for denoising volumetric retinal oct and oct angiography in 3d shearlet domain. Optics Letters 45, 694-697.

Yang, J., Tao, Y., Xu, Q., Zhang, Y., Ma, X., Yuan, S., Chen, Q., 2022. Self-supervised sequence recovery for semi-supervised retinal layer segmentation. IEEE Journal of Biomedical and Health Informatics 26(8), 3872-3883.

Yao, X., Alam, M. N., Le, D., Toslak, D., 2020. Optical coherence tomography. Experimental Biology and Medicine 245(4), 301-312.

Zabel, P., Kaluzny, J. J., Wilkosc-Debczynska, M., Gebska-Toloczko, M., Suwala, K., Zabel, K., Zaron, A., Kucharski, R., Araszkiewicz, A., 2019. Comparison of retinal microvasculature in patients with alzheimers disease and primary open-angle glaucoma by optical coherence tomography angiography. Investigative Ophthalmology & Visual Science 60, 3447-3455.

Zhang, M., Hwang, T. S., Campbell, J. P., Bailey, S. T., Wilson, D. J., Huang, D., Jia, Y., 2016 Projection-resolved optical coherence tomographic angiography. Biomedical Optics Express 7(3), 816-828.

Zhang, Q., Zhang, A., Lee, C. S., Lee, A. Y., Rezaei, K. A., Roisman, L., et al., 2017. Projection artifact removal improves visualization and quantitation of macular neovascularization imaged by optical coherence tomography angiography. Ophthalmology Retina 1, 124-136.

Zhang, Y., Huang, C., Li, M., Xie, S., Xie, K., Yuan, S., Chen, Q., 2020. Robust layer segmentation against complex retinal abnormalities for en face OCTA generation. International Conference on Medical Image Computing and Computer-Assisted Intervention 2020.

Zheng, Y., Gandhi, J. S., Stangos, A. N., Campa, C., Broadbent, D. M., Harding, S. P., 2010. Automated segmentation of foveal avascular zone in fundus fluorescein angiography. Investigative Ophthalmology & Visual Science 51(7), 3653-3659.

Zhou, Z., Siddiquee, M. R., Tajbakhsh, N., Liang, J., 2018. Unet++: a nested u-net architecture for medical image segmentation. International Conference on Medical Image Computing and Computer-Assisted Intervention 2018.

# Supplementary Material

## RELATED WORK

Some related works on vessel segmentation and FAZ segmentation are summarized in Tables S1 and S2.

**Table S1.** Summary of **Vessel** segmentation task on OCTA images.

| | Author | Method | Dataset | Results |
|---|---|---|---|---|
| Thresholding | Terheyden et al., 2020 | 7 thresholding methods: **Manual, Huang, Li, Otsu, Moments, Mean, Percentile** | 15 subjects | - |
| | Mehta et al., 2020 | 11 thresholding methods: **global:** default, Huang, IsoData, mean, Otsu; **local:** Bernsen, mean, median, Niblack, Otsu, Phansalkar | 13 healthy | - |
| Filtering | Chu et al., 2016 | **Frangi filter**, local mean adaptive threshold. | 5 healthy | - |
| | Li et al., 2017 | Top-hat filter enhancement, **optimally oriented flux (OOF).** | 8 images (mice) | DSC=0.848 |
| | Aharony et al., 2019 | **Frangi filter**, Otsu thresholding. | 26 healthy, 20 DR & 6 AMD & 4 RVO | - |
| | Xu et al., 2019 | Otsu thresholding (for large vessel). **Frangi Hessian filter** and global thresholding (for all vessel). | 108 healthy & 123 DR | - |
| Active Contour Models | Eladawi et al., 2017 | GGMRF model for contrast improvement and denoise; **Markov-Gibbs random field** model for segmentation. | 23 healthy & 24 DR | DSC=0.9504 |
| Deep Learning | Prentasic et al., 2016 | **Custom DNN**, pixel-wise classification by square window. | 6 healthy | Accuracy=0.83 |
| | Mou et al., 2019 | **CS-Net & CS²-Net:** U-Net structure with a channel and spatial attention module; vessel skeleton segmentation. | 30 subjects | Accuracy=0.9183 |
| | Stefan et al., 2020 | **3D CNN,** OCTA image enhancement, segmentation and gap-correction. | 2 mice | DSC=0.55 |
| | Pissas et al., 2020 | **iU-Net**: U-Net cascades with shared weights. | 50 subjects | DSC=0.8540 |
| | Lo et al., 2020 | **U-Net structure**, fine-tuning using a transfer learning method. | 8 healthy & 28 DR | DSC(SCP)=0.8599<br>DSC(DVC)=0.7986 |
| | Giarratano et al., 2020 | **U-Net**, **CS-Net**, **Frangi** Filter, **OOF**, etc. | 55 ROIs from 11 subjects | DSC=0.89 |
| | Alam et al., 2020 | **AV-Net:** modified U-Net architecture, **artery-vein segmentation** with OCT/OCTA projection maps input. | 50 images | AV DSC=0.8281 |
| | Li et al., 2020a | **IPN** (projection learning module)**:** Input 3D data and output 2D segmentation. 3D to 2D. | **OCTA-500** | DSC=0.8815 |
| | Ma et al., 2021 | **OCTA-Net:** split-based coarse-to-fine network to detect both thick and thin vessels. | ROSE: 229 images | DSC=0.7576 |
| | Peng et al., 2021 | **FARGO:** a coarse-to-fine cascaded network with spatial attention and channel attention modules | **OCTA-500** | DSC (6 mm) =0.8915<br>DSC (3 mm) =0.9168 |
| | Wu et al., 2021 | **PAENet: IPN based,** add adaptive pooling module and feature fusion module. 3D to 2D. | **OCTA-500** | DSC (6 mm) =0.8969 |
| | Li et al., 2022a | **IMN:** CNN with an up-sampled encoding path and a down-sampled decoding path. | **OCTA-500**, ROSE, Giarratano | DSC=0.9019 |
| | Li et al., 2022c | **RPS-Net: IPN based**, with dual-way projection learning module. 3D to 2D. | **OCTA-500** | DSC (6 mm) =0.8989<br>DSC (3 mm) =0.9155 |
| | Liu et al., 2022 | **ACRROSS:** Semi-supervised segmentation, disentangled representation learning. | **OCTA-500,** ROSE, XJU | DSC (3 mm) =0.912 |

**Table S2.** Summary of **FAZ** segmentation task on OCTA images.

| | Author | Method | Dataset | Results |
|---|---|---|---|---|
| Morphology | Díaz et al., 2019 | White top-hat, **Canny edge detector**, morphological closure, removal of small objects | OCTAGON | Jaccard=0.82 |
| Graph Theory | Xu et al., 2019 | **Graph Cut** | 108 healthy & 123 DR | DSC=0.90 |
| Active Contour Model | Alam et al., 2017 | Active contours without edges | 36 SCR & 26 control | - |
| | Lu et al., 2018 | **GGVF snake model** | 19 healthy & 66 DR | Jaccard>0.82 |
| | Lin et al., 2020 | **Level Sets** macro in ImageJ | 57 healthy | DSC=0.9243 |
| Deep Learning | Guo et al., 2019 | **MEDnet** & **MEDnetv2,** nonperfusion area segmentation. | 76 healthy & 104 DR | - |
| | Guo et al., 2019 | Improved U-Net by appending BN layers and SE blocks. | 45 myopic | DSC=0.976 |
| | Li et al., 2020b | Lightweight UNet-like architecture. | 316 subjects | DSC=0.8468 |
| | Li et al., 2020a | **IPN** (projection learning module)**:** Input 3D data and output 2D segmentation. 3D to 2D. | **OCTA-500** | DSC = 0.8861 |
| | Liang et al., 2021 | Image transform network and UNet-like architecture. | OCTAGON & 45 myopic | DSC=0.9263 & 0.978 |
| | Jabour et al., 2021 | UNet-like architecture with Hausdorff Distance loss. | 204 healthy & DR | DSC=0.909 |
| | Guo et al., 2021 | Customized encoder-decoder network. | 63 healthy & 17 DR | DSC = 0.88 |
| | Mirshahi et al., 2021 | **FPN & Mask R-CNN.** | 37 healthy & 126 DR | DSC = 0.94 |
| | Lin et al., 2021 | **BSDA-Net:** joint FAZ segmentation and multidisease classification. | **OCTA-500**, OCTAGON & FAZID | DSC = 0.9607,0.8864 & 0.9103 |
| | Peng et al., 2021 | **FARGO:** two subnetworks with vessel segmentation as an auxiliary task. | **OCTA-500** | DSC (6 mm) = 0.9272<br>DSC (3 mm) = 0.9839 |
| | Li et al., 2022c | **RPS-Net: IPN based**, with dual-way projection learning module. 3D to 2D. | **OCTA-500** | DSC (6 mm) = 0.9155<br>DSC (3 mm) = 0.9780 |
| | Xu et al., 2022 | **3D U-Net** with non-local attention gate and topological consistency constraint.<br>**3D FAZ segmentation** | **OCTA-500** | DSC (6 mm,3D) = 0.8681<br>DSC (3 mm,3D) = 0.9424 |

## Ablation Study of IPN-V2

The ablation study of IPN-V2 was performed on the OCTA_6mm including the following configurations:
(i) IPN: basic IPN with 4 projection learning modules.
(ii) IPN-V2 with a CNN backbone: the 2D segmentation stage uses a standard CNN without down-sampling and concatenation operations. In this configuration, we aim to measure the segmentation performance in the absence of high-level semantic information.
(iii) IPN-V2 without FPM: the 3D-to-2D projection stage uses a single 1x1 convolution operation with stride h for down-sampling. In this configuration, we aim to illustrate the effectiveness of FPM.
(iv) IPN-V2: the complete IPN-V2 architecture as shown in Fig.11.

The input of the above network is the OCTA volume. Since training the IPN consumes considerable GPU memory, and the entire volume (128 × 400 × 400) cannot be accommodated, the input for the above configurations is unified as a patch (128 × 100 × 100) randomly cropped from the OCTA volume.

The results of these experiments are shown in Table S3. It can be observed that IPN-V2 has a considerable performance improvement compared to IPN. This improvement comes from efficient feature compression by FPM and the high-level semantics provided by U-Net, both of which are indispensable. After adding the high-dimensional semantic information provided by U-Net, the artery segmentation, vein segmentation and FAZ segmentation are significantly improved, which also reflects that the identification of arteries, veins and FAZ is more difficult than that of capillaries.

IPN-V2 greatly reduces GPU memory usage compared to IPN, which means that IPN-V2 allows for a larger size input. We further performed additional ablation studies on the effect of increasing the input size as the following configuration:
(v) IPN-V2: the complete IPN-V2 architecture with an input size of 128 × 200 × 200.
(vi) IPN-V2: the complete IPN-V2 architecture with an input size of 128 × 400 × 400, which means inputting the full 3D volume.

It can be observed that the performance improves as the input size increases, suggesting that larger inputs can provide more global features that are beneficial for segmentation. With model parameters and GPU memory usage less than IPN, IPN-V2 segmented the entire OCTA volume at one time and achieved more than 10% mIoU improvement.

**Table S3.** Results of ablation study of IPN-V2 on OCTA_6mm.

| Methods | Backbone | Input Size (Pixel) | Params (M) | Memory (GB) | IoU (%) | | | | mIoU (%) |
|---|---|---|---|---|---|---|---|---|---|
| | | | | | C | A | V | F | |
| IPN | - | $128*100^2$ | 8.0 | 9.6 | 79.82 | 59.92 | 60.18 | 79.31 | 69.81 |
| IPN-V2 | CNN | $128*100^2$ | 2.3 | 2.1 | 80.60 | 55.28 | 58.55 | 79.21 | 68.41 |
| IPN-V2 w/o FPM | U-Net | $128*100^2$ | 6.5 | 2 | 63.25 | 65.30 | 67.46 | 78.77 | 68.69 |
| IPN-V2 | U-Net | $128*100^2$ | 7.6 | 1.7 | 81.91 | 72.45 | 73.61 | 84.22 | 78.05 |
| IPN-V2 | U-Net | $128*200^2$ | 7.6 | 3.2 | 80.47 | 75.54 | 76.13 | 86.14 | 79.57 |
| IPN-V2 | U-Net | $128*400^2$ | 7.6 | 7.3 | 84.34 | 76.74 | 77.26 | 88.76 | 81.77 |

## SEGMENTATION TASKS ON OCTA-500

We consider each segmentation label provided by OCTA-500 as a separate segmentation task, including large vessel segmentation, capillary segmentation, artery segmentation, vein segmentation, FAZ segmentation, 3D FAZ segmentation, and retinal layer segmentation. We evaluate the performance of different methods on these tasks. For large vessel segmentation, capillary segmentation, artery segmentation, vein segmentation, and FAZ segmentation, we use the baselines, parameter settings and evaluation metrics in this paper; for 3D FAZ segmentation, we use the baselines, parameter settings and evaluation metrics in (Xu et al., 2022); for retinal layer segmentation, we use the baselines, parameter settings and evaluation metrics in (Yang et al., 2022). All experiments are implemented on two subsets of OCTA-500, OCTA_6mm and OCTA_3mm, respectively, using the division of training set, validation set, and test set in Section 5.1. Tables S4-S10 provide quantitative evaluations of the test set. In the tasks of artery segmentation and vein segmentation, the IPN model is difficult to converge during training, so this result is omitted.

Note that, the capillary segmentation, artery segmentation, and vein segmentation here differ from the corresponding subtasks in the CAVF task: the capillaries in the CAVF task exclude arteries and veins; the arteries and veins in the CAVF task intersect. The connections and differences between different tasks can be seen intuitively in Fig. S1, which provides several segmentation examples using IPN-V2 in different tasks. In Fig. S2-S6, we also provide precision-recall curves of large vessel segmentation, capillary segmentation, artery segmentation, vein segmentation, and FAZ segmentation, from which we can more intuitively observe the segmentation performance of different models in these tasks.

**Table S4.** Results (%) of **Large Vessel** segmentation on OCTA-500.

| Methods | OCTA_6mm test set | | | | | OCTA_3mm test set | | | | |
|---|---|---|---|---|---|---|---|---|---|---|
| | Dice | IoU | ACC | SE | SP | Dice | IoU | ACC | SE | SP |
| U-Net | 89.11 | 80.47 | 97.92 | **89.55** | 98.79 | 92.04 | 85.31 | 98.95 | 91.51 | 99.49 |
| UNet ++ | 89.25 | 80.73 | **97.96** | 88.64 | **98.94** | 92.02 | 85.28 | 98.96 | 90.47 | 99.57 |
| UNet 3+ | **89.31** | **80.83** | **97.96** | 88.78 | 98.92 | 91.86 | 85.02 | 98.93 | 91.14 | 99.49 |
| Attention U-Net | 89.26 | 80.74 | 97.95 | 88.93 | 98.89 | 91.91 | 85.10 | 98.95 | 89.93 | **99.60** |
| CS-Net | 87.85 | 78.47 | 97.72 | 86.54 | 98.89 | 90.38 | 82.53 | 98.74 | 89.17 | 99.43 |
| AV-Net | 88.89 | 80.12 | 97.92 | 87.60 | 99.00 | 91.42 | 84.28 | 98.86 | **91.53** | 99.39 |
| IPN | 88.05 | 78.82 | 97.70 | 88.08 | 98.68 | 90.71 | 83.11 | 98.77 | 90.84 | 99.33 |
| IPN-V2 | 89.17 | 80.55 | **97.96** | 88.62 | 98.93 | **92.32** | **85.81** | **99.00** | 91.04 | 99.56 |

**Table S5.** Results (%) of **Capillary** segmentation on OCTA-500.

| Methods | OCTA_6mm test set | | | | | OCTA_3mm test set | | | | |
|---|---|---|---|---|---|---|---|---|---|---|
| | Dice | IoU | ACC | SE | SP | Dice | IoU | ACC | SE | SP |
| U-Net | 88.48 | 79.43 | 91.36 | 89.61 | 92.30 | 90.63 | 82.92 | 92.22 | 91.87 | 92.54 |
| UNet ++ | 88.46 | 79.39 | 91.38 | 89.26 | 92.63 | 90.62 | 82.90 | 92.18 | 92.11 | 92.31 |
| UNet 3+ | 88.48 | 79.42 | 91.37 | 89.58 | 92.40 | 90.63 | 82.93 | 92.19 | 92.17 | 92.26 |
| Attention U-Net | 88.54 | 79.49 | 91.43 | 89.18 | 92.61 | 90.59 | 82.84 | 92.14 | 92.22 | 92.11 |
| CS-Net | 88.16 | 78.90 | 91.13 | 89.27 | 92.14 | 90.17 | 82.14 | 91.67 | 92.99 | 90.74 |
| AV-Net | 87.98 | 78.65 | 90.93 | 90.01 | 91.49 | 89.89 | 81.72 | 91.48 | 92.59 | 90.83 |
| IPN | 93.69 | 88.15 | 95.44 | 92.03 | **97.24** | 95.01 | 90.50 | 95.90 | 95.46 | 96.14 |
| IPN-V2 | **94.82** | **90.17** | **96.18** | **94.94** | 96.73 | **95.55** | **91.49** | **96.36** | **95.65** | **96.82** |

**Table S6.** Results (%) of **Artery** segmentation on OCTA-500.

| Methods | OCTA_6mm test set | | | | | OCTA_3mm test set | | | | |
|---|---|---|---|---|---|---|---|---|---|---|
| | Dice | IoU | ACC | SE | SP | Dice | IoU | ACC | SE | SP |
| U-Net | 87.27 | 77.59 | **98.93** | 85.32 | 99.55 | 89.58 | 81.33 | 99.25 | 88.20 | 99.68 |
| UNet ++ | 87.07 | 77.29 | 98.92 | 84.19 | **99.60** | 89.70 | 81.57 | 99.24 | 90.10 | 99.60 |
| UNet 3+ | 86.66 | 76.66 | 98.87 | 85.68 | 99.48 | 89.65 | 81.40 | 99.24 | 89.44 | 99.62 |
| Attention U-Net | **87.35** | **77.79** | **98.93** | 86.22 | 99.52 | 90.14 | 82.25 | 99.29 | 88.43 | **99.71** |
| CS-Net | 79.89 | 67.22 | 98.33 | 78.54 | 99.24 | 82.12 | 69.98 | 98.73 | 79.76 | 99.46 |
| AV-Net | 84.47 | 73.48 | 98.70 | 82.40 | 99.46 | 78.05 | 64.39 | 98.32 | 80.64 | 99.02 |
| IPN-V2 | 87.27 | 77.56 | 98.92 | **86.29** | 99.49 | **90.78** | **83.23** | **99.33** | **90.26** | 99.67 |

**Table S7.** Results (%) of **Vein** segmentation on OCTA-500.

| Methods | OCTA_6mm test set | | | | | OCTA_3mm test set | | | | |
|---|---|---|---|---|---|---|---|---|---|---|
| | Dice | IoU | ACC | SE | SP | Dice | IoU | ACC | SE | SP |
| U-Net | 86.90 | 77.06 | 98.74 | 85.89 | 99.42 | 87.99 | 78.77 | 99.34 | 84.44 | **99.79** |
| UNet ++ | 87.06 | 77.35 | **98.77** | 84.22 | **99.53** | 88.15 | 79.01 | 99.34 | 86.02 | 99.74 |
| UNet 3+ | 87.24 | 77.58 | 98.75 | 87.38 | 99.35 | 88.12 | 78.98 | 99.34 | 84.83 | 99.78 |
| Attention U-Net | 87.03 | 77.32 | 98.71 | **87.76** | 99.27 | 88.07 | 78.86 | 99.32 | 87.19 | 99.68 |
| CS-Net | 80.48 | 67.78 | 98.03 | 82.44 | 98.84 | 80.91 | 68.21 | 98.93 | 78.31 | 99.56 |
| AV-Net | 84.09 | 72.97 | 98.41 | 84.25 | 99.15 | 69.17 | 53.22 | 98.16 | 72.60 | 98.93 |
| IPN-V2 | **87.30** | **77.67** | **98.77** | 87.62 | 99.34 | **89.71** | **81.46** | **99.42** | **88.31** | 99.75 |

**Table S8.** Results (%) of **FAZ** segmentation on OCTA-500.

| Methods | OCTA_6mm test set | | | | | OCTA_3mm test set | | | | |
|---|---|---|---|---|---|---|---|---|---|---|
| | Dice | IoU | ACC | SE | SP | Dice | IoU | ACC | SE | SP |
| U-Net | 87.55 | 81.36 | 99.73 | 86.25 | **99.92** | 96.48 | 93.42 | 99.73 | 97.24 | 99.83 |
| UNet ++ | 84.10 | 76.87 | 99.68 | 84.61 | 99.89 | 97.15 | 94.65 | 99.79 | 97.25 | 99.90 |
| UNet 3+ | 87.28 | 80.39 | 99.68 | 89.53 | 99.85 | 96.94 | 94.21 | 99.77 | 96.87 | 99.89 |
| Attention U-Net | 86.73 | 80.25 | 99.71 | 85.51 | 99.93 | **97.76** | **95.68** | **99.83** | **97.78** | **99.92** |
| CS-Net | 86.16 | 78.03 | 99.66 | 85.83 | 99.87 | 95.48 | 91.74 | 99.65 | 95.33 | 99.84 |
| AV-Net | 81.44 | 72.57 | 99.59 | 82.89 | 99.83 | 96.97 | 94.26 | 99.77 | 97.09 | 99.89 |
| IPN | 84.14 | 74.11 | 99.66 | 79.82 | **99.92** | 94.02 | 89.44 | 99.55 | 93.08 | 99.83 |
| IPN-V2 | **90.12** | **83.45** | **99.76** | **92.18** | 99.88 | 97.68 | 95.56 | **99.83** | 97.45 | **99.92** |

**Table S9.** Results of **3D FAZ** segmentation on OCTA-500.

| Methods | OCTA_6mm test set | | OCTA_3mm test set | |
|---|---|---|---|---|
| | Dice (%) ↑ | HD95 (vox) ↓ | Dice (%) ↑ | HD95 (vox) ↓ |
| 3D UNet | 92.37 ± 10.05 | 2.39 ± 3.67 | 96.50 ± 3.59 | 1.64 ± 3.62 |
| VNet | 91.31 ± 9.96 | 2.85 ± 4.52 | 96.28 ± 2.35 | 1.75 ± 3.05 |
| Attention U-Net | 92.63 ± 9.57 | 2.45 ± 3.85 | 96.97 ± 2.05 | **1.12 ± 0.47** |
| UNet++ | **93.60 ± 7.64** | 2.01 ± 2.78 | 96.56 ± 2.75 | 1.99 ± 4.79 |
| DeepLabV3+ | 92.24 ± 8.42 | **1.67 ± 2.22** | 95.97 ± 3.99 | 1.52 ± 2.07 |
| CANet | 92.90 ± 6.02 | 2.61 ± 3.82 | 96.67 ± 3.59 | 1.35 ± 1.60 |
| Xu et al., 2022 | 93.34 ± 9.05 | 2.06 ± 3.27 | **97.39 ± 2.19** | 1.13 ± 0.59 |

**Table S10.** Results (Dice) of **Layer** segmentation on OCTA-500.

| Methods | OCTA_6mm test set | | | | | OCTA_3mm test set | | | | |
|---|---|---|---|---|---|---|---|---|---|---|
| | ILM<br>IPL | IPL<br>OPL | OPL<br>ISOS | ISO<br>RPE | RPE<br>BM | ILM<br>IPL | IPL<br>OPL | OPL<br>ISOS | ISO<br>RPE | RPE<br>BM |
| U-Net | 97.33 | 92.59 | 94.38 | 93.30 | 87.75 | 98.38 | 94.93 | 97.17 | **96.69** | **92.86** |
| BRUnet | 97.48 | 92.22 | 94.17 | 92.47 | **87.91** | 98.41 | 95.15 | 97.24 | 96.55 | 92.46 |
| RelayNet | 97.48 | 92.35 | 93.31 | 91.85 | 86.23 | 98.27 | 94.65 | 97.00 | 96.40 | 92.31 |
| UNet++ | 97.47 | 92.24 | 93.65 | 91.77 | 86.18 | 98.23 | 94.58 | 96.94 | 96.31 | 92.29 |
| DeepLabV3+ | **97.93** | **93.85** | **94.69** | **93.38** | 87.05 | 97.98 | 94.47 | 96.81 | 96.13 | 91.46 |
| Yang et al., 2022 | 97.78 | 93.15 | **94.69** | 93.22 | 86.18 | **98.55** | **95.18** | **97.54** | 96.02 | 92.46 |

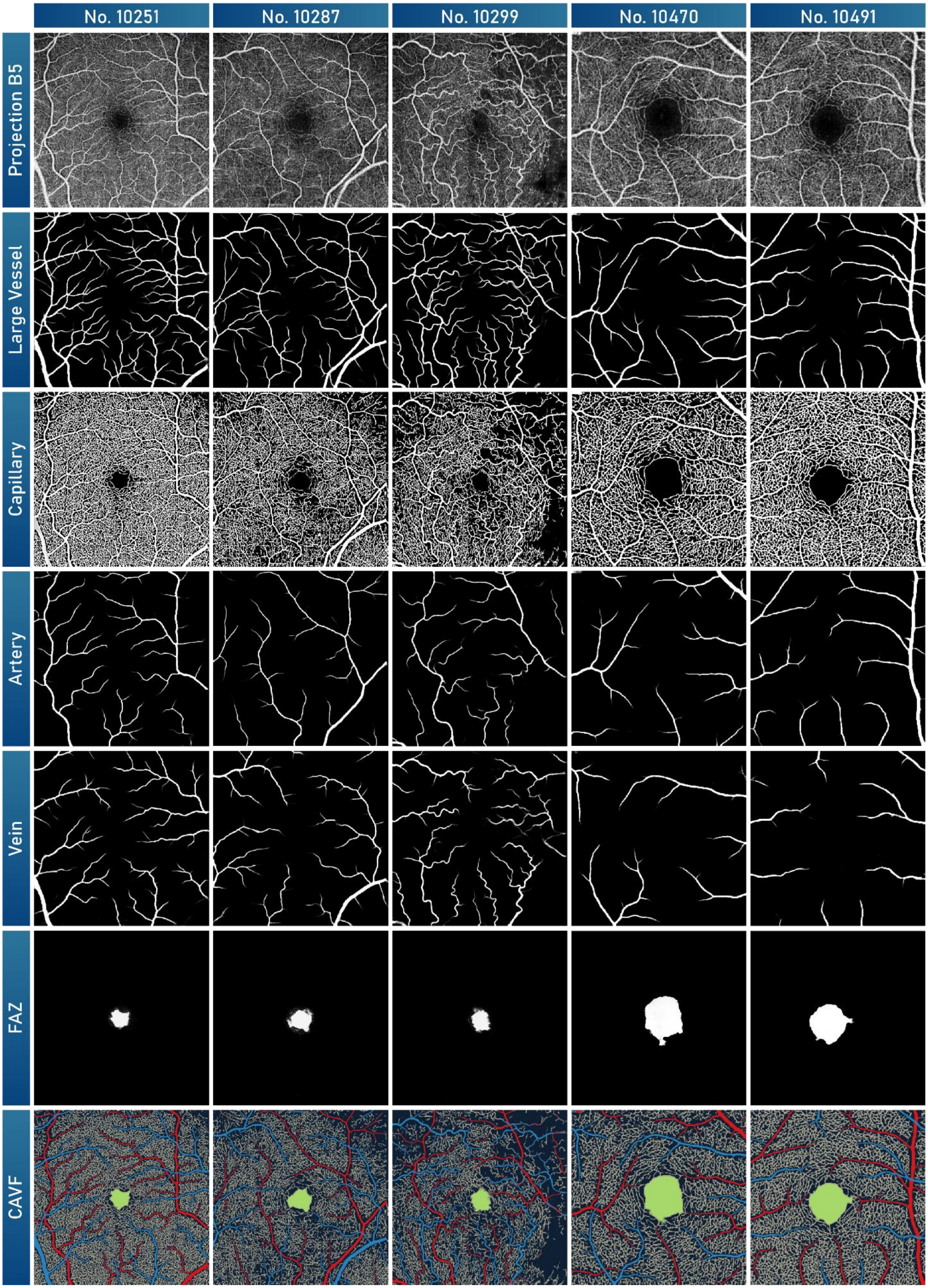

**Fig. S1.** Examples of segmentation results using IPN-V2 in different tasks.

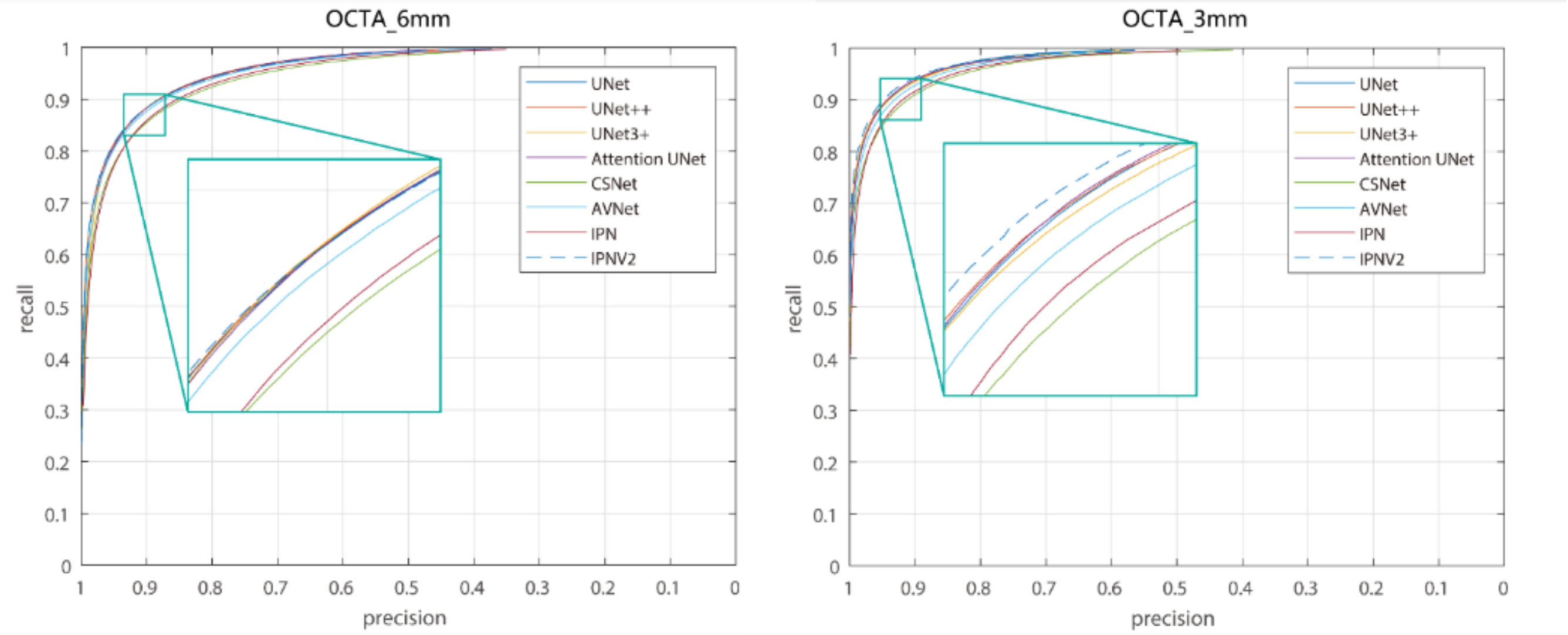

**Fig. S2.** PR curves of different methods in the **Large Vessel** segmentation.

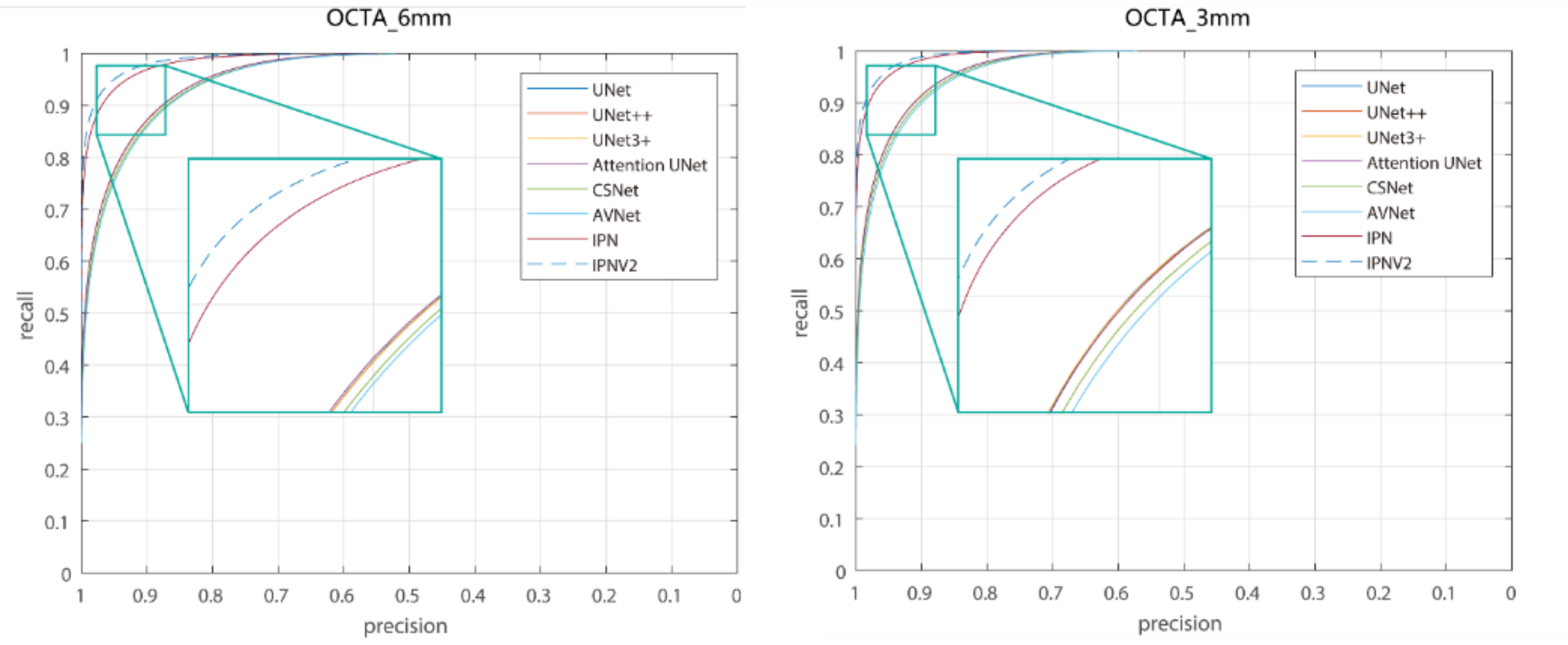

**Fig. S3.** PR curves of different methods in the **Capillary** segmentation.

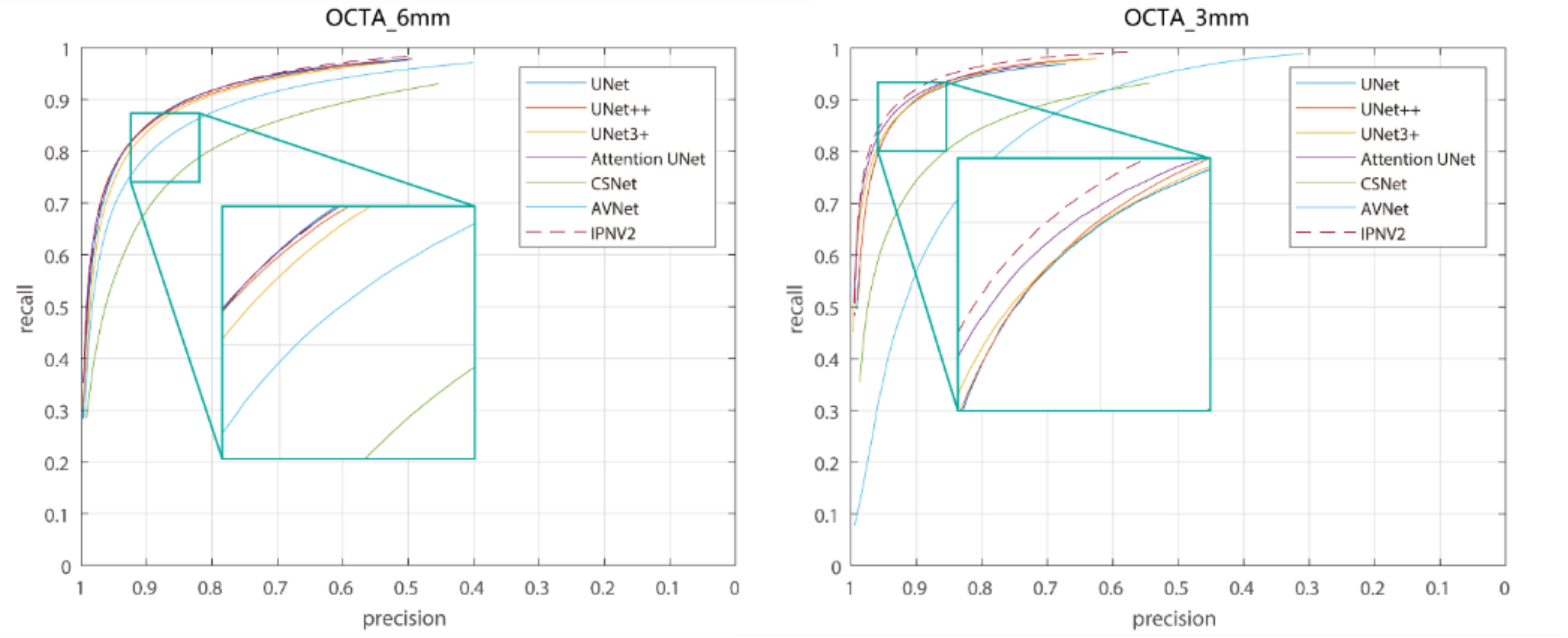

**Fig. S4.** PR curves of different methods in the **Artery** segmentation.

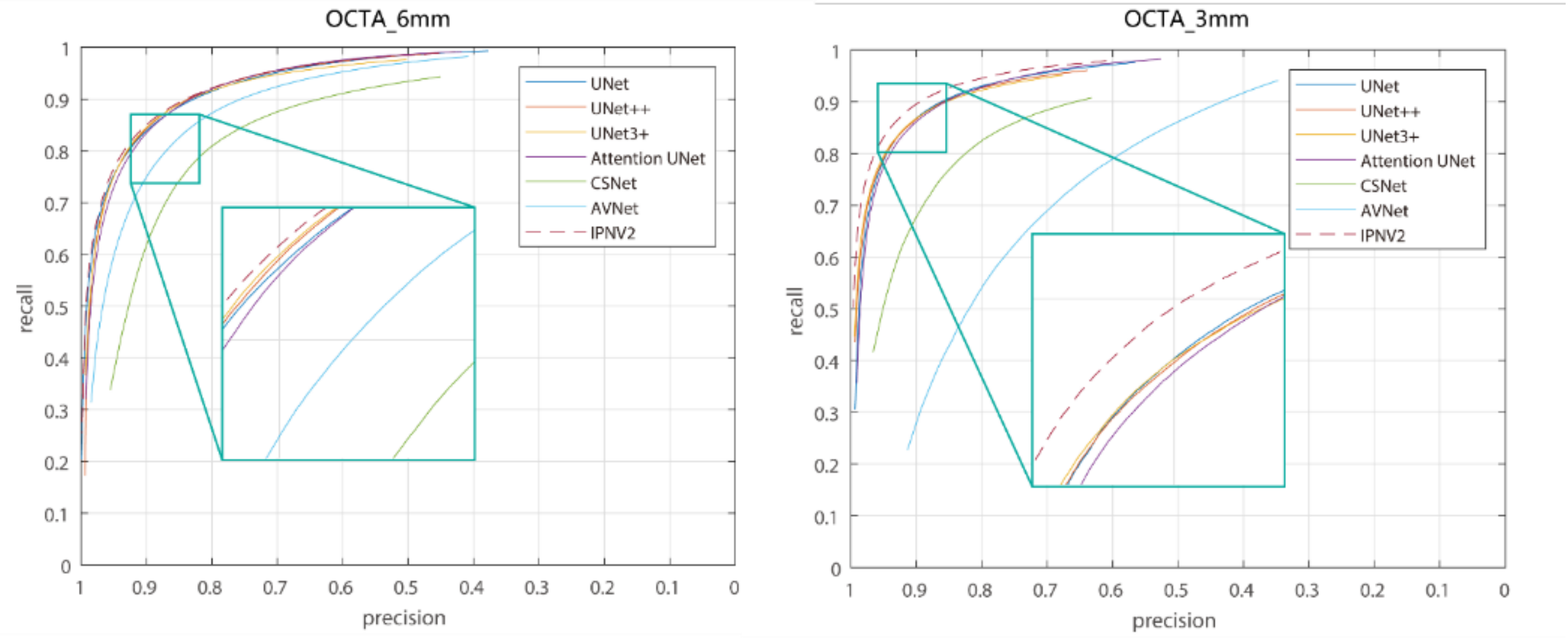

**Fig. S5.** PR curves of different methods in the **Vein** segmentation.

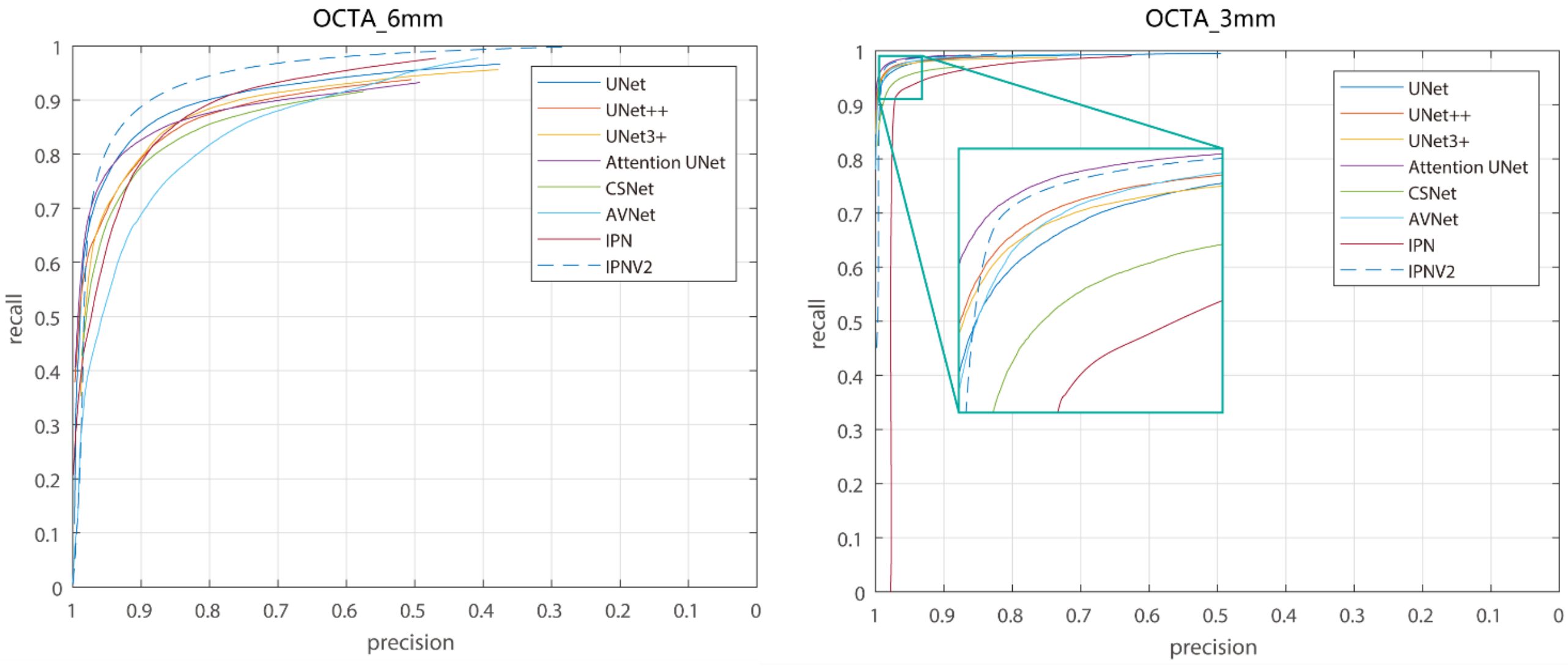

**Fig. S6.** PR curves of different methods in the **FAZ** segmentation.

## EVALUATION OF COMPARATIVE EXPERIMENTS

To facilitate researchers to query more evaluation indicators, we present the evaluation of Dice, accuracy, sensitivity and specificity in the CAVF task in Tables S11-S14. Table 15 provides the speed evaluation of different baselines. The 2D-to-2D baselines use less GPU memory for training and show faster testing than the 3D-to-2D baselines because a 3D-to-2D network takes more time to read the 3D volumes, and convolution on a 3D volume also requires more memory. Nonetheless, benefiting from our optimization, IPN-V2 has achieved test speeds close to 2D baselines. IPN-V2 completed the test with at least 2 subjects in 1 second, which is an acceptable speed. Note that, the 2D baselines take the projection maps as input, and the generation of the projection maps relies on layer segmentation. If the time for layer segmentation is considered, the 2D baselines will be slower than IPN-V2.

**Table S11.** Results (Dice, %) of **CAVF** task on OCTA-500.

| Methods | OCTA_6mm test set | | | | | OCTA_3mm test set | | | | |
|---|---|---|---|---|---|---|---|---|---|---|
| | C | A | V | F | Mean | C | A | V | F | Mean |
| U-Net | 82.90 | 86.20 | 87.06 | 91.83 | 86.99 | 87.97 | 88.08 | 87.71 | 97.51 | 90.32 |
| UNet ++ | 82.98 | 86.26 | **87.30** | 91.56 | 87.03 | 87.97 | 89.28 | 88.39 | 97.60 | 90.81 |
| UNet 3+ | 82.92 | 85.72 | 86.79 | 89.30 | 86.18 | 87.92 | 89.28 | 88.43 | 97.25 | 90.72 |
| Attention U-Net | 83.22 | 86.31 | 87.20 | 90.95 | 86.92 | 88.07 | 89.61 | 88.79 | **97.83** | 91.08 |
| CS-Net | 81.54 | 78.11 | 80.33 | 86.30 | 81.57 | 86.93 | 83.53 | 81.20 | 96.02 | 86.92 |
| AV-Net | 82.51 | 84.61 | 85.38 | 88.44 | 85.23 | 87.49 | 81.56 | 76.95 | 96.96 | 85.74 |
| IPN | 88.69 | 74.61 | 74.69 | 87.31 | 81.32 | 91.04 | 82.97 | 80.63 | 95.24 | 87.47 |
| IPN-V2 | **91.46** | **86.73** | 87.06 | **93.62** | **89.72** | **92.53** | **90.17** | **89.65** | 97.47 | **92.46** |

**Table S12.** Results (Accuracy, %) of **CAVF** task on OCTA-500.

| Methods | OCTA_6mm test set | | | | | OCTA_3mm test set | | | | |
|---|---|---|---|---|---|---|---|---|---|---|
| | C | A | V | F | Mean | C | A | V | F | Mean |
| U-Net | 91.07 | 98.88 | **98.76** | 99.80 | 97.12 | 92.03 | 99.17 | 99.29 | 99.81 | 97.57 |
| UNet ++ | 91.09 | 98.87 | **98.76** | 99.78 | 97.13 | 91.87 | 99.24 | 99.35 | 99.81 | 97.57 |
| UNet 3+ | 90.95 | 98.84 | 98.70 | 99.77 | 97.06 | 91.99 | 99.22 | 99.33 | 99.79 | 97.58 |
| Attention U-Net | 91.11 | 98.88 | **98.76** | 99.77 | 97.13 | 92.06 | 99.25 | 99.37 | **99.83** | 97.63 |
| CS-Net | 90.19 | 98.20 | 98.09 | 99.60 | 96.52 | 91.05 | 98.79 | 98.94 | 99.70 | 97.12 |
| AV-Net | 90.83 | 98.73 | 98.58 | 99.68 | 96.95 | 91.69 | 98.65 | 98.74 | 99.76 | 97.21 |
| IPN | 94.33 | 97.79 | 97.60 | 99.66 | 97.35 | 94.21 | 98.79 | 98.89 | 99.66 | 97.89 |
| IPN-V2 | **96.65** | **98.89** | 98.74 | **99.84** | **98.28** | **95.11** | **99.29** | **99.41** | 99.80 | **98.40** |

**Table S13.** Results (Sensitivity, %) of **CAVF** task on OCTA-500.

| Methods | OCTA_6mm test set | | | | | OCTA_3mm test set | | | | |
|---|---|---|---|---|---|---|---|---|---|---|
| | C | A | V | F | Mean | C | A | V | F | Mean |
| U-Net | 84.12 | 84.00 | 85.22 | 91.84 | 86.29 | 89.50 | 84.85 | 87.44 | 97.75 | 89.89 |
| UNet ++ | 84.33 | 84.99 | 86.62 | **93.69** | 87.41 | 91.30 | 87.28 | 84.92 | **98.11** | 90.40 |
| UNet 3+ | 85.53 | 83.02 | 86.65 | 88.31 | 85.88 | 89.51 | 89.30 | 88.48 | 97.59 | 91.22 |
| Attention U-Net | 85.85 | 84.01 | 85.71 | 92.29 | 86.96 | 90.00 | **89.80** | 86.74 | **98.11** | 91.16 |
| CS-Net | 84.50 | 77.20 | 78.43 | 85.95 | 81.52 | 91.50 | 83.89 | 79.67 | 95.94 | 87.75 |
| AV-Net | 84.28 | 83.79 | 84.13 | 90.39 | 85.65 | 89.19 | 81.34 | 74.85 | 96.85 | 85.56 |
| IPN | 86.92 | 77.21 | 72.80 | 92.74 | 82.42 | 90.99 | 80.22 | 80.08 | 93.85 | 86.29 |
| IPN-V2 | **90.80** | **86.65** | **87.20** | 93.60 | **89.56** | **93.53** | 89.15 | **88.96** | 97.98 | **92.40** |

**Table S14.** Results (Specificity, %) of **CAVF** task on OCTA-500.

| Methods | OCTA_6mm test set | | | | | OCTA_3mm test set | | | | |
|---|---|---|---|---|---|---|---|---|---|---|
| | C | A | V | F | Mean | C | A | V | F | Mean |
| U-Net | 93.51 | 99.53 | **99.46** | 99.93 | 98.11 | 93.27 | **99.71** | 99.64 | 99.90 | 98.13 |
| UNet ++ | 93.40 | 99.49 | 99.39 | 99.89 | 98.04 | 92.17 | 99.69 | **99.79** | 99.89 | 97.88 |
| UNet 3+ | 92.83 | **99.55** | 99.32 | 99.93 | 97.91 | 93.22 | 99.60 | 99.65 | 99.88 | 98.09 |
| Attention U-Net | 92.92 | 99.54 | 99.44 | 99.90 | 97.96 | 93.09 | 99.61 | 99.75 | 99.90 | 98.09 |
| CS-Net | 92.11 | 99.13 | 99.10 | 99.80 | 97.54 | 90.82 | 99.36 | 99.52 | 99.85 | 97.39 |
| AV-Net | 93.15 | 99.39 | 99.32 | 99.83 | 97.92 | 92.95 | 99.32 | 99.46 | 99.88 | 97.90 |
| IPN | 96.76 | 98.68 | 98.88 | 99.77 | 98.52 | 95.64 | 99.50 | 99.44 | **99.91** | 98.62 |
| IPN-V2 | **97.19** | 99.43 | 99.34 | **99.94** | **98.97** | **95.87** | 99.67 | 99.72 | 99.87 | **98.78** |

**Table S15.** Speed and space evaluation of different baselines on OCTA_6mm.

| Methods | GPU | Resolution | Memory (GB) | Params | Speed (ms) |
|---|---|---|---|---|---|
| CS-Net | RTX 3090 | 400 × 400 | 1.0 | 8.4 M | 144 |
| AV-Net | RTX 3090 | 400 × 400 | 2.4 | 6.5 M | 158 |
| U-Net | RTX 3090 | 400 × 400 | 2.0 | 5.5 M | 238 |
| UNet ++ | RTX 3090 | 400 × 400 | 5.3 | 12.1 M | 255 |
| UNet 3+ | RTX 3090 | 400 × 400 | 4.9 | 16.1 M | 243 |
| Attention U-Net | RTX 3090 | 400 × 400 | 2.8 | 5.6 M | 197 |
| IPN | RTX 3090 | 128 × 100 × 100 | 9.6 | 8.0 M | 2336 |
| IPN-V2 | RTX 3090 | 128 × 400 × 400 | 7.3 | 7.6 M | 427 |